\documentclass[12pt]{article}

\usepackage[dvipsnames]{xcolor}
\usepackage{float}
\usepackage{graphicx,wrapfig,lipsum}
\usepackage{subfigure}
\usepackage{eurosym}
\usepackage{amsmath}
\usepackage{natbib}
\usepackage{amssymb}
\usepackage{amsfonts}
\usepackage{setspace}
\usepackage{color}
\usepackage{graphicx}
\usepackage{fancyhdr}

\usepackage[all]{nowidow}
\usepackage[splitrule,flushmargin,bottom,symbol*]{footmisc}
\usepackage[colorlinks,citecolor=blue,urlcolor=blue]{hyperref}
\usepackage{bm}
\usepackage{enumitem,amssymb}
\usepackage{rotating}
\usepackage{tcolorbox}
\usepackage{tabularx}
\usepackage{booktabs}
\usepackage[english]{babel}
\usepackage[flushleft]{threeparttable}
\usepackage[letterpaper,top=2cm,bottom=2cm,left=3cm,right=3cm,marginparwidth=1.75cm]{geometry}
\usepackage{adjustbox}
\usepackage{mathtools}

\RequirePackage[OT1]{fontenc}
\RequirePackage{hypernat}

\newtheorem{theorem}{Theorem}
\newtheorem{definition}{Definition}

\newtheorem{proposition}{Proposition}

\newtheorem{lemma}{Lemma}
\newtheorem{remark}{Remark}

\newenvironment{proof}{\emph{Proof.} }{\mbox{ } \hfill $\blacksquare$ \vspace{2mm}}

\graphicspath{{figures/}}

\begin{document}

\title{Concave Rationalization with an Ideal Point: An Afriat Theorem and an Application to Survey Design\thanks{I am grateful to the editor, the associate editor, and three anonymous referees for detailed and constructive comments that substantially improved the paper. I also thank Thomas Baudin, Alberto Bisin, Daniel Chen, Christina Felfe, Garance G\'enicot, Jun Hyung Kim, Mathieu Lefebvre, Olivier L'Haridon, Judith Saurer, Patrick Schneider, Thierry Verdier, and Romain Warcziarg for their insights and encouragements, and the seminar audiences at the SAET annual meeting, Lund University, and Aix Marseille School of Economics. Mistakes are mine. I acknowledge funding from the French government under the ANR JCJC “BIAS” project (reference: ANR-25-CE26-
7164-01) and from the ``France 2030'' investment plan managed by the French National Research Agency (reference: ANR-17-EURE-0020) and from Excellence Initiative of Aix-Marseille University - A*MIDEX.}}

\singlespacing \author{{\large Avner Seror\thanks{avner.seror@univ-amu.fr; Aix Marseille Univ, CNRS, AMSE, Marseille, France}}}

\date{July 2026}

\maketitle

\begin{abstract}
\noindent
This paper develops an Afriat-type characterization of concave rationalization with an unknown ideal point. We show that, for each candidate peak, a finite system of linear inequalities is necessary and sufficient for the existence of a continuous concave utility with an ideal point that rationalizes choices from linear budget sets anchored at different corners of the choice space. A stronger characterization adds the requirement that supergradients at observed choices point coordinatewise toward the peak, a necessary condition for single-peaked rationalizability. The resulting peak-oriented system has a transparent geometry - budgets anchored at different corners triangulate the ideal point - and yields a nonparametric set of candidate ideal points. This provides the theoretical foundation for the Priced Survey Methodology (PSM), in which respondents complete the same survey under different linear constraints. We apply the PSM to study political preferences in a sample of French respondents.\\

\end{abstract}

\renewcommand{\thefootnote}{\arabic{footnote}}
\onehalfspacing
\setcounter{page}{1}

\section{Introduction}\label{section: intro}

Single-peaked preferences are a foundational concept across economics. An agent has single-peaked preferences if there is an ideal point - a most-preferred alternative - and preferences decline as outcomes move away from it. In social choice theory, single-peakedness guarantees the existence of a Condorcet winner and the strategy-proofness of the median rule \citep{black48, sen1970, moulin1980}; in political economy it underpins the median voter theorem and spatial models of electoral competition. Single-peaked preferences may also arise naturally in survey responses. When a respondent is asked how much she agrees with a statement on a numerical scale, it is natural to posit an ideal answer - the number that best expresses her view - with agreement declining as the recorded answer moves away from it. If this is the right model of how internal states map into survey responses, then the object of interest is the respondent's ideal answer, and the analyst's task is twofold: to check whether the model holds in the first place, and to recover the ideal answer when it does. This paper develops the revealed-preference foundations for both tasks.

A single unconstrained answer on a numerical scale reveals little about a respondent's underlying preferences. The \emph{Priced Survey Methodology} (PSM) elicits revealed-preference information by asking the same questions repeatedly under varying linear constraints. In each round, each question is assigned a ``price'' and the respondent must report answers whose total cost equals a fixed budget; the prices encode the trade-off, so that raising the answer to one question forces a proportional lowering of the answer to another. She cannot report her ideal on every question at once, and the trade-off she makes - how she allocates agreement across questions as prices and budgets vary - reveals how she values moving each answer toward its ideal. Where a single unconstrained answer is a number on a scale with no built-in analytic structure, the revealed preferences elicited from constrained choices give analytic content to the notion of an ``ideal answer'': it is the maximizer, on the answer space, of a utility function consistent with the observed trade-offs.

Because the ideal answer may lie anywhere on the scale, budgets anchored at a single corner reveal trade-offs from only one direction. The PSM therefore anchors budget sets at different corners of the answer space, so that budgets from different corners jointly \emph{triangulate} the candidate ideal point. This corner-anchoring is the source of identification, and it is what distinguishes the PSM from prior experiments that recover preferences from choices on linear budget sets \citep{andreoni2002, syngjoo2007, choi2014_rationality, fisman2015_science, halevy2018}.

Our main theoretical contribution is an Afriat-type characterization of when such constrained choices are consistent with a concave utility that has an ideal point. \citeauthor{afriat1967}'s classical theorem states that finite choice data from linear budgets are rationalizable by a concave, increasing utility if and only if a finite system of linear inequalities - the Afriat inequalities - is feasible. We extend this to ideal-point preferences over corner-anchored budgets. Our first theorem shows that the data are rationalizable by a continuous concave utility with an ideal point if and only if an augmented Afriat system is feasible, in which the unknown peak enters as a virtual observation carrying the highest utility. Without a monotonicity restriction, however, this system is permissive. The constant utility satisfies it, so the theorem is an existence result rather than a test. Our second theorem supplies the empirical content: it adds the requirement that the supergradient at each observed choice point coordinatewise toward the peak (a \emph{peak-oriented} sign restriction) and have a strictly positive projection toward the peak whenever the observation differs from it. These restrictions capture a testable, finite-data implication of single-peakedness - the orientation of the supergradient at observed choices - and they are the ideal-point counterpart of the non-satiation restriction that gives Afriat's theorem its empirical bite. 

Additionally, the peak-oriented system has a transparent dual structure. Geometrically, each round restricts the candidate peak to a ``double cone'' around the observed choice. Behaviorally, once the orientation is fixed by a candidate peak, and away from candidate peaks that coincide with retained observations, the data are jointly rationalizable if and only if the reoriented signed normals satisfy the usual GARP cycle condition - an oriented analog of Afriat's behavioral primitive. This yields a nonparametric \emph{peak set} of candidate ideal points that the corner-anchored budgets triangulate without any parametric assumption.

The nonparametric peak set localizes the ideal answer to a region without delivering a unique point. To measure how well a respondent's choices conform to peak-oriented rationalization, we adopt a Houtman--Maks consistency index - the largest fraction of rounds that can be jointly rationalized at a common ideal point. The peak set is then defined precisely: the candidate ideal answers that attain this maximum. The consistency index is the test promised at the outset: it asks whether the data are compatible with any single-peaked mapping from internal states to recorded answers in the first place.

We apply the PSM to political preferences in a sample of 469 French respondents who answered two questions about policy priorities, first without constraints and then under 19 different linear constraints. The corner-anchored budgets triangulate the ideal answer tightly: the median respondent's peak set covers less than $1\%$ of the answer space, and the average consistency index across subjects is $0.82$. These findings rest on the constrained choices alone; the unconstrained stated answer enters only as a calibration input that positions the linear constraints, and not in the Afriat inequalities themselves. A parametric single-peaked specification provides complementary point estimates of the ideal answer and the relative importance the respondent assigns to each question.

This paper contributes to three literatures. First, it contributes to the literature that generalizes and applies \citeauthor{afriat1967}'s seminal theorem \citep{matzkin91, forges2009, polisson2013, reny2015, nishimura2017}.\footnote{Proofs of \citeauthor{afriat1967}'s theorem can be found in \cite{afriat1967}, \cite{varian1982}, \cite{fostel_scarf_todd2004}, \cite{chambers_echenique_2016}, and \cite{polisson2016}. \cite{DZIEWULSKI2024103016} provides a comprehensive overview.} The paper provides a finite revealed-preference characterization of concave rationalization with an unknown ideal point, for budget sets anchored at different corners of the choice space. Any single-peaked rationalization satisfies these peak-oriented conditions.\footnote{\cite{ok_2015} develops a revealed preference theory of reference-dependent choice behavior, which is related but distinct.}

Second, it contributes to the economic literature on survey methodology, an increasingly important area of research \citep{Stantcheva2022, dacunto2024}. Interpretation and aggregation of survey responses raise several challenges - subjects perceive scales differently \citep{bond2019}, and cognitive factors affect how they answer \citep{tanur1992questions, sudman96} - and a recent literature explores corrections, including open-ended questions \citep{stancheva2022_aerpp}, scale-use adjustments through multiple surveys \citep{benjamin2023}, response error models \citep{bertrand2001}, scoring methods \citep{prelec2004_science}, and limited self-knowledge \citep{falk2021}. The PSM provides an alternative: it elicits preferences from the pattern of constrained choices, and provides an internal test of whether those choices admit a single-peaked structure in the first place. The closest theoretical comparison is \citet{apesteguia2023}, whose theory of survey rationalizability uses endorsement data on a common one-dimensional scale; our framework uses constrained numerical choices in a multidimensional setting and exploits the trade-offs respondents make across questions.

Third, it contributes to the literature on single-peaked preferences \citep{sen1970, bossert2009, ballester2011, lackner2017, chatterji2016, PUPPE2018}. Our definition of single-peakedness on a multidimensional answer space is the natural analog of \citet{black48}'s single-peakedness from each corner of the choice space, and on a discrete grid coincides with the notion of \citet{barbera1993}.

The paper proceeds as follows. Section~\ref{section: theory} presents the PSM design, the two main characterization results, and the link to single-peaked preferences. Section~\ref{section: rationality} defines the consistency index and the nonparametric peak set. Section~\ref{section: application} presents the empirical application. Section~\ref{section: discussion} concludes.

\section{Design and Theory}\label{section: theory}

\subsection{Notations and Setup}

Consider a survey with $S$ questions. Let $\mathcal{S}=\{1,\dots, S\}$ denote the set of questions and $\mathcal{R}=\{1,\dots, R\}$ the set of rounds. For each question $s \in \mathcal{S}$, the set of possible answers is $X(s) = [0, M_s]$ for some $M_s > 0$. The set of possible answer vectors is $X=\prod_{s\in \mathcal{S}} X(s)$, a compact hyperrectangle in $\mathbb{R}^S$. Let $\mathcal{C}(X)$ denote the set of corners (vertices) of $X$: there are $2^S$ corners. For example, if $S=2$ and $X=[0,10]^2$, then $\mathcal{C}(X)=\{(0,0), (10,0), (0,10), (10,10)\}$.

For any $\mathbf{x} \in X$ and corner $c \in \mathcal{C}(X)$, we write $\mathbf{x}_c$ for the vector $\mathbf{x}$ expressed in the coordinate system with origin $c$. Concretely, if $c = (c_1, \ldots, c_S)$, then $x_{s,c} = |x_s - c_s|$ for each $s \in \mathcal{S}$. For instance, if $X = [0,10]^2$ and a respondent answers $(2,7)$, then $\mathbf{x}_{(0,0)}=(2,7)$, $\mathbf{x}_{(10,0)}=(8,7)$, $\mathbf{x}_{(0,10)}=(2,3)$, and $\mathbf{x}_{(10,10)}=(8,3)$. To simplify notation, we omit the corner subscript when $c$ is the origin, writing $\mathbf{x}$ for $\mathbf{x}_{(0,\ldots,0)}$.

Let $\mathcal{I}=\{1,\dots, I\}$ denote the set of participants. For participant $i \in \mathcal{I}$, let $q_s^{r,i}$ be the answer to question $s$ in round $r$. We drop the participant index to simplify notation. The set of observations is $X^{o}=\{\mathbf{q}^r\}_{r\in \mathcal{R}}\subset X$, and $\mathcal{B}^r \subseteq X$ denotes the choice set in round $r$.

\subsection{Choice Sets}\label{subsection: choice sets}

In each round $r \in \mathcal{R}$, the respondent faces a linear budget constraint:
\begin{equation}\label{eq: budget sets}
    \mathcal{B}^r=\{\mathbf{q}\in X:\mathbf{q_{o^r}\cdot p^r}= m^r\},
\end{equation}
where $o^r\in \mathcal{C}(X)$ is the corner associated with round $r$, $\mathbf{p}^r\in \mathbb{R}_+^S$ is a ``price'' vector, and $m^r > 0$ is the budget level. The choice set $\mathcal{B}^r$ is a hyperplane - a downward-sloping ``budget line'' in the coordinate system with origin $o^r$. When $o^r$ is the standard origin, this reduces to a classical budget constraint. The key departure from the standard setting is that budget lines are anchored at different corners across rounds, so the respondent faces trade-offs from multiple directions.

\begin{figure}[!htbp]
    \centering
    \subfigure{\includegraphics[width=0.3\textwidth]{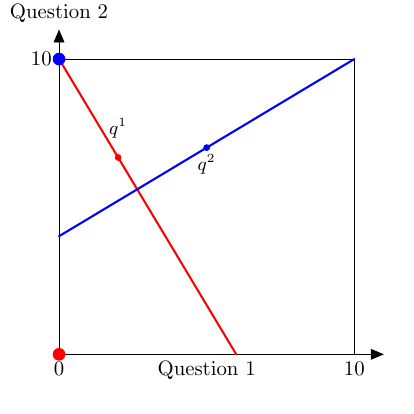}}
    \caption{Example with two rounds}
    \label{fig:fig1}
\end{figure}

It is convenient to express each budget set in standard coordinates. Define the signed price vector $\mathbf{a}^r \in \mathbb{R}^S$ by
\begin{equation}\label{eq: signed prices}
a_s^r = \begin{cases} p_s^r & \text{if } o_s^r = 0, \\ -p_s^r & \text{if } o_s^r = M_s, \end{cases}
\end{equation}
and let $\mu^r = m^r - \sum_{s: o_s^r = M_s} p_s^r M_s$. Then $\mathcal{B}^r = \{\mathbf{x} \in X : \mathbf{a}^r \cdot \mathbf{x} = \mu^r\}$. Figure~\ref{fig:fig1} illustrates budget sets anchored at two different corners.

\medskip
\noindent\textbf{Standing assumption (nondegenerate budget).} Throughout the paper we assume that every budget is nondegenerate, in the sense that
\[
0 < m^r < \sum_{s=1}^S p^r_s\, M_s \qquad \text{for every } r \in \mathcal{R}.
\]
Equivalently, each hyperplane $\mathcal{B}^r$ meets the interior of $X$ rather than only a face or corner.\footnote{This is the constraint qualification invoked in the first-order condition of Lemma~\ref{lem:normal-supergradient} and used throughout the appendix. The experimental design in Section~\ref{section: application} satisfies the assumption.}

\medskip
\noindent\textbf{Observed feasibility.} We further assume $\mathbf{q}^r \in \mathcal{B}^r$ for every $r \in \mathcal{R}$: each observed choice lies on the round-$r$ budget. The survey interface enforces this through co-dependent sliders that keep the displayed answer on the budget line.

\subsection{Single-Peaked Preferences}

Let $D=\{\mathbf{q}^r, \mathcal{B}^r\}_{r\in \mathcal{R}}$ denote an individual-level dataset.

\paragraph{Standing price condition.} Throughout the theoretical analysis we maintain the \emph{strict-price} condition: $p_s^r > 0$ for every $r \in \mathcal{R}$ and $s = 1,\ldots,S$, so that $a_s^r \neq 0$ for every coordinate.

We do not assume that observed choices lie in the interior of $X$. Survey respondents often anchor at the scale endpoints (we report below that $35\%$ of constrained-round observations sit on the boundary of $X$). Under the standing budget-interiority assumption, the first-order condition for maximizing a continuous concave $u$ on $\mathcal{B}^r$ delivers a supergradient parallel to the budget normal at every observed choice, interior or boundary (Lemma~\ref{lem:normal-supergradient} in Appendix~\ref{appendix: proofs}):
\[
  \lambda^r \mathbf{a}^r \in \partial u(\mathbf{q}^r) \qquad \text{for some } \lambda^r \in \mathbb{R}.
\]

\begin{definition}[Rationalization]\label{def: rationalizes}
A utility function $u: X\rightarrow \mathbb{R}$ \emph{rationalizes} $D$ if $\mathbf{q}^r \in \arg\max_{\mathbf{x} \in \mathcal{B}^r} u(\mathbf{x})$ for all $r \in \mathcal{R}$.
\end{definition}

\begin{definition}[Single-peaked]\label{def: single peaked}
A function $f:X\rightarrow \mathbb{R}$ is \emph{single-peaked} if:
\begin{enumerate}
    \item There exists a unique $\mathbf{y}^*\in X$ such that $f(\mathbf{y})\leq f(\mathbf{y}^*)$ for all $\mathbf{y}\in X$.
    \item For any $\mathbf{x}, \mathbf{y} \in X$ such that $\mathbf{x}_c\leq \mathbf{y}_c\leq \mathbf{y}^*_c$ for some $c \in \mathcal{C}(X)$, we have $f(\mathbf{x})\leq f(\mathbf{y})$.
\end{enumerate}
\end{definition}

The second condition states that if $\mathbf{y}$ lies between $\mathbf{x}$ and the peak $\mathbf{y}^*$ in some coordinate system, then $\mathbf{y}$ is weakly preferred to $\mathbf{x}$: moving toward the peak always weakly increases utility. This is the multivariate analog of \citet{black48}'s single-peakedness, formulated as monotonicity in the product order from the corner relative to which $\mathbf{y}$ lies between $\mathbf{x}$ and $\mathbf{y}^*$; on a discrete grid it coincides with the multidimensional single-peakedness of \citet{barbera1993}, characterized via $L_1$-betweenness.

\begin{definition}[Single-peaked rationalization]\label{def: sp rationalization}
$D$ admits a \emph{single-peaked rationalization} if there exists a continuous concave single-peaked utility $u$ that rationalizes $D$.
\end{definition}

\subsection{Main Result}

\begin{definition}[Concave rationalization with an ideal point]\label{def: concave ideal}
$D$ admits a \emph{concave rationalization with an ideal point} if there exist a point $\mathbf{y}^* \in X$ and a continuous concave utility function $u: X \to \mathbb{R}$ such that $\mathbf{y}^* \in \arg\max_{\mathbf{x} \in X} u(\mathbf{x})$ and $\mathbf{q}^r \in \arg\max_{\mathbf{x} \in \mathcal{B}^r} u(\mathbf{x})$ for all $r \in \mathcal{R}$.
\end{definition}

\begin{theorem}\label{theorem: halevy}
The following three statements are equivalent.
\begin{enumerate}
\item[(i)] $D$ admits a concave rationalization with an ideal point (Definition~\ref{def: concave ideal}).

\item[(ii)] There exist $\mathbf{y}^* \in X$, numbers $U^0, U^1, \ldots, U^R \in \mathbb{R}$, and scalars $\lambda^1, \ldots, \lambda^R \in \mathbb{R}$ such that, writing $\mathbf{g}^r := \lambda^r \mathbf{a}^r$, the following hold for all $r, l \in \mathcal{R}$:
\begin{align}
U^r &\leq U^l + \mathbf{g}^l \cdot (\mathbf{q}^r - \mathbf{q}^l), \label{eq: afriat1}\\
U^r &\leq U^0, \label{eq: afriat2}\\
U^0 &\leq U^l + \mathbf{g}^l \cdot (\mathbf{y}^* - \mathbf{q}^l). \label{eq: afriat3}
\end{align}

\item[(iii)] There exist $\mathbf{y}^* \in X$, numbers $U^0, U^1, \ldots, U^R$, and scalars $\lambda^r \in \mathbb{R}$ such that, writing $\mathbf{g}^r := \lambda^r \mathbf{a}^r$, the function
\begin{equation}\label{eq: utility}
u(\mathbf{x}) := \min\bigl\{U^0,\; \min_{l \in \mathcal{R}} \bigl[ U^l + \mathbf{g}^l \cdot (\mathbf{x} - \mathbf{q}^l) \bigr] \bigr\}, \qquad \mathbf{x} \in X,
\end{equation}
is continuous and concave, satisfies $u(\mathbf{q}^r) = U^r$ for all $r$, $u(\mathbf{y}^*) = U^0 = \max_{\mathbf{x} \in X} u(\mathbf{x})$, and rationalizes the data.
\end{enumerate}
\end{theorem}

The system~\eqref{eq: afriat1}--\eqref{eq: afriat3} is the Afriat system for an unknown ideal point. Inequality~\eqref{eq: afriat1} is the supergradient inequality of $u$ at $\mathbf{q}^l$ evaluated at $\mathbf{q}^r$; inequality~\eqref{eq: afriat2} states that the ideal point has the highest utility; inequality~\eqref{eq: afriat3} states that the supergradient at each observation supports the peak. The relation $\mathbf{g}^l = \lambda^l \mathbf{a}^l$ encodes the first-order condition for maximizing a concave $u$ on $\mathcal{B}^l$, which is available under the standing budget-interiority assumption (Lemma~\ref{lem:normal-supergradient}). The peak $\mathbf{y}^*$ is part of the system; however, the system is very permissive, as the following remark shows.

\begin{remark}[Degeneracy]\label{rem: degeneracy}
Theorem~\ref{theorem: halevy} is an exact characterization, but it is empirically weak. Setting $\lambda^r = 0$ (hence $\mathbf{g}^r = \mathbf{0}$) and $U^r = U^0$ for all $r$ makes the system~\eqref{eq: afriat1}--\eqref{eq: afriat3} feasible for any dataset and any candidate peak $\mathbf{y}^*$: every inequality reduces to $U^0 \le U^0$. The associated utility is the constant function $u \equiv U^0$. Unlike the classical Afriat theorem - where monotonicity excludes the constant utility and gives the system its empirical bite - the ideal-point setting drops monotonicity (the peak is interior), so the constant solution is admitted here. Theorem~\ref{theorem: halevy} should therefore be read as a broad existence result rather than as a testable restriction on the data. The next subsection develops the orientation and nondegeneracy restrictions that exclude this trivial solution.
\end{remark}

\subsection{A Stronger Characterization: Peak-Oriented Rationalization}

Theorem~\ref{theorem: halevy} characterizes concave rationalization with an ideal point, but since it allows the trivial constant-utility solution, it does not provide a meaningful empirical test. To obtain a stronger and testable characterization, we impose two additional restrictions: supergradients must point coordinatewise toward the ideal point, and they must have a strictly positive projection on the displacement to the peak at observations that differ from it.

\begin{definition}[Peak-oriented concave rationalization]\label{def: peak oriented}
$D$ admits a \emph{peak-oriented concave rationalization} if there exist $\mathbf{y}^* \in X$, a continuous concave $u: X \to \mathbb{R}$, and scalars $\lambda^r \in \mathbb{R}$ for each $r \in \mathcal{R}$ such that, writing $\mathbf{g}^r := \lambda^r \mathbf{a}^r$:
\begin{enumerate}
\item[(a)] $\mathbf{y}^* \in \arg\max_{\mathbf{x} \in X} u(\mathbf{x})$;
\item[(b)] $\mathbf{q}^r \in \arg\max_{\mathbf{x} \in \mathcal{B}^r} u(\mathbf{x})$ for every $r \in \mathcal{R}$;
\item[(c)] $\mathbf{g}^r \in \partial u(\mathbf{q}^r)$ for every $r \in \mathcal{R}$;
\item[(d)] $g^r_s\,(y^*_s - q^r_s) \geq 0$ for every $r \in \mathcal{R}$ and $s = 1, \ldots, S$;
\item[(e)] if $\mathbf{q}^r \neq \mathbf{y}^*$, then $\mathbf{g}^r \cdot (\mathbf{y}^* - \mathbf{q}^r) > 0$.
\end{enumerate}
\end{definition}

Condition (c) requires that the supergradient at each observed choice be parallel to the budget normal $\mathbf a^r$; this is the first-order condition for a maximizer of a concave $u$ on the budget hyperplane, available under the standing budget-interiority assumption (Lemma~\ref{lem:normal-supergradient}). The supergradient set $\partial u(\mathbf{q}^r)$ is the collection of all vectors $\mathbf{g}$ satisfying $u(\mathbf{x}) \leq u(\mathbf{q}^r) + \mathbf{g} \cdot (\mathbf{x} - \mathbf{q}^r)$ for all $\mathbf{x} \in X$, and is nonempty at each observed choice by Lemma~\ref{lem:normal-supergradient} (given condition (b) and the nondegenerate-budget assumption).

Condition (d) adds a directional restriction: the supergradient at each observed choice must point coordinatewise toward the peak $\mathbf{y}^*$. Concretely, if $y^*_s > q^r_s$ (the peak is above the choice in dimension $s$), then $g^r_s \geq 0$; if $y^*_s < q^r_s$, then $g^r_s \leq 0$. Since $\mathbf{g}^r = \lambda^r \mathbf{a}^r$, this fixes the sign of $\lambda^r$ once the candidate peak is fixed.

Condition (e) is a strict directional nondegeneracy requirement: whenever the observed choice differs from the peak, the selected supergradient must have a strictly positive projection on the displacement toward the peak. Combined with (d), this is equivalent to requiring at least one coordinate — among those in which $\mathbf{q}^r$ and $\mathbf{y}^*$ differ — to carry a strictly oriented supergradient component. This rules out the trivial constant-utility rationalization identified in Remark~\ref{rem: degeneracy} whenever at least one retained observation differs from the candidate peak (see Remark~\ref{rem: constant excluded} for a proof), and is necessary under single-peakedness: strict preference for the peak together with the supergradient inequality gives $\mathbf{g}^r \cdot (\mathbf{y}^* - \mathbf{q}^r) \ge u(\mathbf{y}^*) - u(\mathbf{q}^r) > 0$.

\begin{theorem}[Peak-oriented Afriat theorem]\label{thm: peak oriented}
The following are equivalent.
\begin{enumerate}
\item[(i)] $D$ admits a peak-oriented concave rationalization.
\item[(ii)] There exist $\mathbf{y}^* \in X$, $U^0, U^1, \ldots, U^R \in \mathbb{R}$, and scalars $\lambda^1, \ldots, \lambda^R \in \mathbb{R}$ such that, writing $\mathbf{g}^r := \lambda^r \mathbf{a}^r$, the inequalities~\eqref{eq: afriat1}--\eqref{eq: afriat3} hold, and, for all $r \in \mathcal{R}$ and $s = 1, \ldots, S$:
\begin{equation}\label{eq:po_afriat_sign}
g^r_s (y^*_s - q^r_s) \geq 0,
\end{equation}
and, in addition,
\begin{equation}\label{eq:po_nondegen}
\mathbf{q}^r \neq \mathbf{y}^* \;\Longrightarrow\; \mathbf{g}^r \cdot (\mathbf{y}^* - \mathbf{q}^r) > 0 \qquad \forall\, r \in \mathcal{R}.
\end{equation}
\item[(iii)] There exist $\mathbf{y}^* \in X$, numbers $U^0, U^1, \ldots, U^R$, and scalars $\lambda^1, \ldots, \lambda^R \in \mathbb{R}$ such that, writing $\mathbf{g}^r := \lambda^r \mathbf{a}^r$, the function~\eqref{eq: utility} satisfies all properties in Theorem~\ref{theorem: halevy}(iii), each $\mathbf{g}^r$ belongs to $\partial u(\mathbf{q}^r)$, and conditions~\eqref{eq:po_afriat_sign} and~\eqref{eq:po_nondegen} hold.
\end{enumerate}
\end{theorem}

Condition~\eqref{eq:po_afriat_sign} is the coordinatewise orientation: the supergradient $\mathbf{g}^r = \lambda^r \mathbf a^r$ at each observed choice must point toward the peak. With $\mathbf{g}^r$ parallel to $\mathbf a^r$, orientation reduces to $\lambda^r a^r_s (y^*_s - q^r_s) \geq 0$; if the sign products $a^r_s(y^*_s - q^r_s)$ have opposite signs across dimensions for some round $r$ and candidate peak $\mathbf{y}^*$, no nonzero $\lambda^r$ can satisfy orientation and the round is inconsistent with peak-oriented rationalization at $\mathbf{y}^*$. Condition~\eqref{eq:po_nondegen} is a strict directional nondegeneracy requirement: the selected supergradient has strictly positive projection on the displacement from the observation toward the peak. Under the strict-price condition, orientation, and $\mathbf{q}^r \ne \mathbf{y}^*$, condition~\eqref{eq:po_nondegen} is equivalent to $\lambda^r \ne 0$ (in fact, $\lambda^r$ has the sign making at least one term $\lambda^r a^r_s(y^*_s - q^r_s)$ strictly positive at a coordinate where $\mathbf{q}^r$ and $\mathbf{y}^*$ differ). Together, \eqref{eq:po_afriat_sign} and~\eqref{eq:po_nondegen} give the theorem empirical content: whenever at least one retained observation differs from the candidate peak, they exclude both the constant-utility rationalization and any supergradient that carries no directional information about the peak.


\begin{proposition}\label{prop: sp implies po}
If $D$ admits a single-peaked rationalization (Definition~\ref{def: sp rationalization}), then $D$ admits a peak-oriented concave rationalization (Definition~\ref{def: peak oriented}). In particular, the system~\eqref{eq: afriat1}--\eqref{eq:po_nondegen} is feasible.
\end{proposition}

\begin{remark}[Intermediate class]\label{rem: intermediate class}
Theorem~\ref{thm: peak oriented} characterizes an intermediate class: stronger than unrestricted concave rationalization with an ideal point (Theorem~\ref{theorem: halevy}), and weaker than global single-peakedness. The sign and nondegeneracy conditions are local restrictions on the supergradient at observed choices, not on all points of the domain, and the utility may be non-strictly concave away from the observations.
\end{remark}

\paragraph{Three nested classes.}
The three rationalization concepts we have defined form a hierarchy at the dataset level:
\[
\begin{aligned}
&\text{single-peaked rationalization (Def.~\ref{def: sp rationalization})}\\
\Rightarrow\ &\text{peak-oriented concave rationalization (Def.~\ref{def: peak oriented})}\\
\Rightarrow\ &\text{concave rationalization with an ideal point (Def.~\ref{def: concave ideal}).}
\end{aligned}
\]
The second implication is immediate: a peak-oriented concave rationalization is, in particular, a concave rationalization with an ideal point (drop conditions (c)--(e)). The first is Proposition~\ref{prop: sp implies po} above. The distinctions concern how utility moves toward the peak. A concave utility with an ideal point is automatically monotone toward $\mathbf{y}^*$ along straight segments, since a concave function whose maximum on a line is attained at an endpoint is monotone on that line. Single-peakedness (Definition~\ref{def: single peaked}) is strictly stronger: it requires monotonicity toward $\mathbf{y}^*$ in the coordinatewise order from every corner, not only along straight rays. Peak-oriented rationalization (Definition~\ref{def: peak oriented}) is intermediate: it imposes the coordinatewise orientation of the supergradient \emph{only at observed choices}.

\begin{figure}[!htbp]
    \centering
    \includegraphics[width=0.95\textwidth]{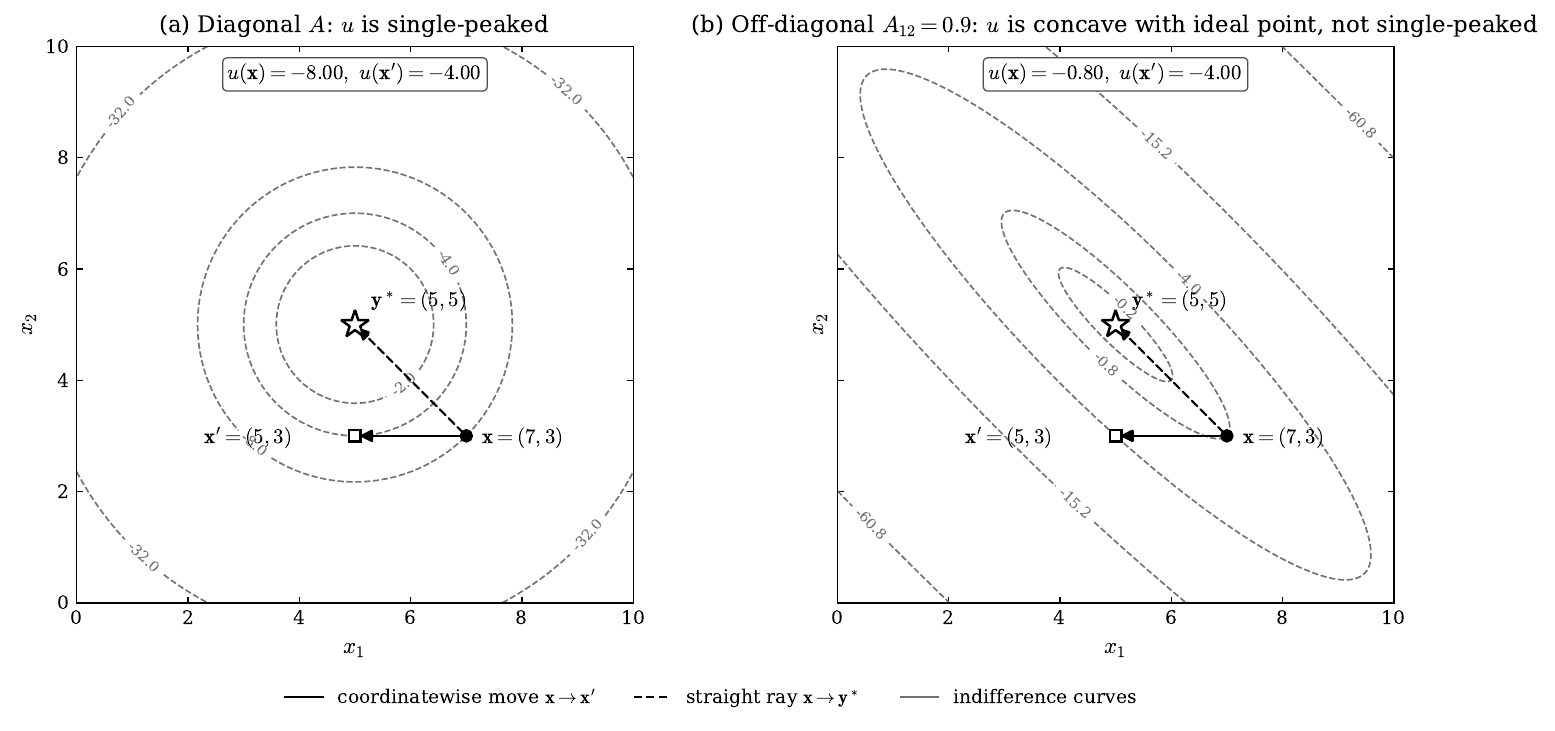}
    \caption{The three rationalization classes are nested but distinct. Both panels plot indifference curves of the concave quadratic $u(\mathbf{x}) = -(\mathbf{x}-\mathbf{y}^*)^\top A(\mathbf{x}-\mathbf{y}^*)$ with peak $\mathbf{y}^*=(5,5)$ and consider the coordinatewise move from $\mathbf{x}=(7,3)$ to $\mathbf{x}'=(5,3)$. Panel~(a): $A$ diagonal - the move toward $\mathbf{y}^*$ in $x_1$ raises utility, and $u$ is single-peaked (Definition~\ref{def: single peaked}). Panel~(b): $A$ has off-diagonal entry $A_{12}=0.9$ - the same coordinatewise move \emph{lowers} utility, so $u$ is concave with an ideal point but not single-peaked. The dashed ray $\mathbf{x}\to\mathbf{y}^*$ is monotone in both panels, illustrating that concavity guarantees monotonicity along straight segments; only single-peakedness adds coordinatewise monotonicity.}
    \label{fig: tilted ellipse}
\end{figure}

The gap between the three classes is genuine. Figure~\ref{fig: tilted ellipse} illustrates a concave quadratic $u(\mathbf{x}) = -(\mathbf{x}-\mathbf{y}^*)^\top A(\mathbf{x}-\mathbf{y}^*)$ with $A$ positive definite: when $A$ is diagonal, $u$ is single-peaked; when $A$ has off-diagonal terms, $u$ remains concave with an ideal point but moving toward $\mathbf{y}^*$ in one coordinate can lower utility, so single-peakedness fails\footnote{Panel~(b) of Figure~\ref{fig: tilted ellipse} has a unique peak with nested concave indifference curves around it and might informally read as single-peaked. We take the stronger coordinatewise view in Definition~\ref{def: single peaked} because that is precisely the property the peak-oriented sign condition tests.} and the peak-oriented sign condition~\eqref{eq:po_afriat_sign} can fail at some choices. This is why finite revealed-preference data cannot pin down single-peakedness directly: concave rationalization with an ideal point is always available (Remark~\ref{rem: degeneracy}); single-peakedness is a global property; and the peak-oriented conditions are exactly the part of single-peakedness that observed choices can reveal - the orientation of the supergradient at each $\mathbf{q}^r$.

\subsection{Behavioral and Geometric Content of Theorem~\ref{thm: peak oriented}}\label{subsection: geometric content}

Theorem~\ref{thm: peak oriented} admits two complementary readings. Geometrically, each round's sign condition restricts the candidate peak to a ``double cone'' around the observed choice (Lemma~\ref{lem: double cone}). Behaviorally, once the orientation is fixed by a candidate peak, and away from candidate peaks that coincide with retained observations, peak-oriented rationalization at the candidate is equivalent to the usual GARP cycle condition applied to the reoriented signed normals (Proposition~\ref{prop: reduction} and Lemma~\ref{lem: garp reduction}). Together these turn the theorem into an operational two-step procedure: a coordinate-by-coordinate cone-membership test, then a signed-normal GARP check on the reoriented data. Section~\ref{section: rationality} uses this two-step procedure to define and compute the consistency index and peak set. Throughout we maintain the standing assumption that $a^r_s\neq 0$ for all $r,s$ (equivalently, strictly positive prices), so that every coordinate is constrained.

\paragraph{The double cone of a round.}
Fix a round $r$ and define the two opposite orthant-cones with apex $\mathbf{q}^r$,
\[
  O^r_{+}:=\{\mathbf{x}\in X: a^r_s(x_s-q^r_s)\ge 0\ \ \forall s\},
  \qquad
  O^r_{-}:=\{\mathbf{x}\in X: a^r_s(x_s-q^r_s)\le 0\ \ \forall s\},
\]
and let $C^r:=O^r_{+}\cup O^r_{-}$ be the \emph{double cone} of round $r$.

\begin{lemma}[Orientation]\label{lem: double cone}
Under the standing assumption $a^r_s\neq 0$, fix a round $r$ and a candidate peak $\mathbf{y}\neq\mathbf{q}^r$. There exists $\lambda^r\neq 0$ satisfying the orientation $\lambda^r a^r_s(y_s-q^r_s)\ge 0$ for all $s$ if and only if $\mathbf{y}\in C^r$; any $\lambda^r>0$ works when $\mathbf{y}\in O^r_{+}$ and any $\lambda^r<0$ works when $\mathbf{y}\in O^r_{-}$, so the sign of $\lambda^r$ is determined by $\mathbf{y}$. Moreover $O^r_{+}\cap O^r_{-}=\{\mathbf{q}^r\}$, and for $S\ge 2$ and $\mathbf{q}^r \in \mathrm{int}(X)$ the set $C^r$ is not convex.
\end{lemma}

Lemma~\ref{lem: double cone} is the geometric content of~\eqref{eq:po_afriat_sign}: each constrained choice $\mathbf{q}^r$ ``votes'' that the ideal point lies in one of the two coordinatewise directions away from $\mathbf{q}^r$ permitted by the signed price $\mathbf{a}^r$, and a peak is admissible for round $r$ only if it lies in that double cone.

\paragraph{Fixed-peak reduction to an oriented Afriat system.}
With the orientation pinned down by the candidate peak, the rest of Theorem~\ref{thm: peak oriented} reduces to a single ordinary Afriat system. For each $r$ with $\mathbf{y}\in C^r$, let $\sigma_r(\mathbf{y})\in\{+1,-1\}$ be the orientation of Lemma~\ref{lem: double cone} and set $\mathbf{b}^r(\mathbf{y}):=\sigma_r(\mathbf{y})\,\mathbf{a}^r$, the budget normal reoriented toward the candidate peak.

\begin{definition}[Peak-oriented rationalizable at a candidate peak]\label{def: peak-oriented at y}
A subset $T\subseteq\mathcal{R}$ is \emph{peak-oriented rationalizable at $\mathbf{y}\in X$} if there exist $U^0$, $\{U^r\}_{r\in T}$, and $\{\lambda^r\}_{r\in T}$ satisfying~\eqref{eq: afriat1}--\eqref{eq:po_nondegen} on $\{\mathbf{q}^r,\mathcal{B}^r\}_{r\in T}$ with the peak held fixed at $\mathbf{y}$.
\end{definition}

\begin{proposition}[Fixed-peak reduction]\label{prop: reduction}
Fix $\mathbf{y}\in X$ and a subset $T\subseteq\mathcal{R}$ such that $\mathbf{q}^r\neq\mathbf{y}$ for all $r\in T$. Then $T$ is peak-oriented rationalizable at $\mathbf{y}$ if and only if
\begin{enumerate}
\item[(a)] $\mathbf{y}\in\bigcap_{r\in T}C^r$; and
\item[(b)] the \emph{augmented oriented Afriat system} - with observations $(\mathbf{x}^r,\mathbf{b}^r(\mathbf{y}))=(\mathbf{q}^r,\sigma_r(\mathbf{y})\mathbf{a}^r)$ for $r\in T$ and virtual top observation $(\mathbf{x}^0,\mathbf{b}^0)=(\mathbf{y},\mathbf{0})$ - is feasible, i.e.\ there exist $U^0,\{U^r\}_{r\in T}$ and $\{\eta^r>0\}_{r\in T}$ such that
\[
  U^j\le U^i+\eta^i\,\mathbf{b}^i(\mathbf{y})\cdot(\mathbf{x}^j-\mathbf{x}^i),
  \qquad i\in T,\ j\in T\cup\{0\},
\]
and
\[
  U^r\le U^0,\qquad r\in T.
\]
\end{enumerate}
\end{proposition}

Proposition~\ref{prop: reduction} reduces peak-oriented rationalizability at $\mathbf{y}$ to two ingredients: a coordinate-by-coordinate cone-membership test ($\mathbf{y}\in C^r$ for every retained $r$) and an Afriat feasibility check on the data with budget normals reoriented toward $\mathbf{y}$. The next step gives the second ingredient a behavioral reading.

\paragraph{Oriented GARP and the behavioral primitive of Theorem~\ref{thm: peak oriented}.}

\begin{definition}[Augmented oriented GARP]\label{def: oriented garp}
Fix $\mathbf{y}\in X$ and $T\subseteq\mathcal{R}$ with $\mathbf{y}\in C^r$ and $\mathbf{y} \neq \mathbf{q}^r$ for every $r\in T$. The augmented oriented data $\{(\mathbf{q}^r,\mathbf{b}^r(\mathbf{y}))\}_{r\in T}\cup\{(\mathbf{y},\mathbf{0})\}$ satisfy \emph{augmented oriented GARP at $\mathbf{y}$} if there is no cycle $i_1,\ldots,i_K,i_{K+1}=i_1$ in $T \cup \{0\}$ - with index $0$ denoting the virtual peak observation ($\mathbf{x}^0 = \mathbf{y}$, $\mathbf{b}^0 = \mathbf{0}$) and indices in $T$ denoting observed choices ($\mathbf{x}^r = \mathbf{q}^r$) - such that $\mathbf{b}^{i_k} \cdot (\mathbf{x}^{i_{k+1}} - \mathbf{x}^{i_k}) \le 0$ for all $k = 1,\ldots,K$, with at least one strict inequality.
\end{definition}

\begin{lemma}[Reduction to retained-round GARP]\label{lem: garp reduction}
Fix $\mathbf{y}\in X$ and $T\subseteq\mathcal{R}$ with $\mathbf{y}\in C^r$ and $\mathbf{y} \neq \mathbf{q}^r$ for every $r\in T$. Under the standing assumption $a^r_s \neq 0$ for all $r,s$, the augmented oriented data satisfy augmented oriented GARP at $\mathbf{y}$ if and only if the retained oriented data $\{(\mathbf{q}^r, \mathbf{b}^r(\mathbf{y}))\}_{r \in T}$ satisfy GARP - that is, there is no cycle $r_1,\ldots,r_K,r_{K+1}=r_1$ in $T$ such that $\mathbf{b}^{r_k}(\mathbf{y})\cdot(\mathbf{q}^{r_{k+1}}-\mathbf{q}^{r_k})\le 0$ for all $k$, with at least one strict inequality.
\end{lemma}

By the Afriat--GARP equivalence for signed normals (Lemma~\ref{lem: afriat-garp signed} in Appendix~\ref{appendix: proofs}), the augmented oriented Afriat system of Proposition~\ref{prop: reduction}(b) is feasible if and only if the augmented oriented data satisfy augmented oriented GARP at $\mathbf{y}$. Combining this with Lemma~\ref{lem: garp reduction}, the system is feasible if and only if the retained oriented data $\{(\mathbf{q}^r, \mathbf{b}^r(\mathbf{y}))\}_{r \in T}$ satisfy the algebraic GARP cycle condition. This identification gives Theorem~\ref{thm: peak oriented} a behavioral primitive that plays a role parallel to the role classical GARP plays in Afriat's theorem. In the classical setup, an observation reveals preference over all bundles affordable at the observed prices, and rationalization is equivalent to acyclicity of that revealed-preference relation. Here, an observation first orients its budget normal toward a candidate peak; once oriented, the observation reveals preference in the oriented half-space, and rationalization at $\mathbf{y}$ (away from coincident peaks) is equivalent to acyclicity of the oriented revealed-preference relation.

Oriented GARP plays the role of behavioral primitive for Theorem~\ref{thm: peak oriented}, parallel to the role classical GARP plays for Afriat's theorem; the only novelty is that the orientation depends on the unknown peak. Because the orientation is a sign vector $\sigma(\mathbf{y})\in\{+1,-1\}^R$ constant on each ``sign cell'' of peak space (Proposition~\ref{prop: cells}), the search over $\mathbf{y}\in X$ is a search over finitely many orientation patterns. Section~\ref{section: rationality} carries it out, producing the consistency index $\kappa^*(D)$ - the largest fraction of rounds jointly rationalizable at some peak - and the nonparametric peak set $\mathcal{Y}^*(D)$, the set of ideal points at which $\kappa^*$ is attained.

\subsection{Summary}

This section has established the theoretical core of the paper. Theorem~\ref{theorem: halevy} gives an exact Afriat-type representation: choices from corner-anchored budget sets are rationalizable by a continuous concave utility with an ideal point if and only if the Afriat system~\eqref{eq: afriat1}--\eqref{eq: afriat3} is feasible. The representation is permissive: the constant utility satisfies it (Remark~\ref{rem: degeneracy}), reflecting the absence of monotonicity rather than a failure of the classical Afriat theorem, which excludes constants through non-satiation. Theorem~\ref{thm: peak oriented} supplies the empirical content by adding the peak-oriented sign condition and the nondegeneracy condition; together these require the supergradient at each observed choice to point coordinatewise toward the ideal point. These conditions capture the local, observed-choice orientation that single-peakedness implies and provide a testable finite-data implication.

The intuition has both a geometric and a behavioral reading. Geometrically, a choice on a budget line anchored at a given corner reveals, coordinate by coordinate, the direction in which the respondent's candidate peak must lie relative to that choice; as corners and prices vary across rounds, these directional restrictions accumulate, and a peak-oriented rationalization exists exactly when they can be reconciled with a single common candidate peak. Behaviorally, at candidate peaks that do not coincide with a retained observation, this reconciliation is acyclicity of the oriented revealed-preference relation - oriented GARP - which plays the role of classical GARP for peak-oriented rationalization. The corner-anchored budgets jointly ``triangulate'' the peak set in both senses. Section~\ref{section: rationality} makes this reconciliation precise, defines a consistency index that measures how much of the data is reconcilable, and characterizes the set of candidate peaks the data can support.

\section{Measuring Consistency}\label{section: rationality}

Theorem~\ref{thm: peak oriented} provides a testable characterization of peak-oriented concave rationalization. The sign and nondegeneracy conditions~\eqref{eq:po_afriat_sign}--\eqref{eq:po_nondegen}, combined with the Afriat inequalities, impose non-trivial restrictions on the data. When a dataset violates these restrictions, it is natural to measure how far it is from conforming to them. Several indices of approximate rationality serve this purpose, including the Critical Cost Efficiency Index \citep{afriat1972}, Varian's index \citep{varian1990}, the money-pump index \citep{echenique2011}, the index of \cite{houtman_maks} which records the largest fraction of observations that can be jointly rationalized, and the minimum cost index of \citet{dean2016}.

We adopt the Houtman--Maks index, for a reason specific to our setting. The Critical Cost Efficiency Index is constructed by radially deflating each budget toward a fixed origin - a construction that requires monotonicity and a single natural deflation point. Single-peakedness has no monotone analog (deviating from the ideal answer in either direction reduces utility), and the PSM's budgets are anchored at different corners across rounds, with no single deflation point that respects all corners uniformly. Adapting the Critical Cost Efficiency Index to a corner-anchored setting would require methodological choices - which deflation point, whether the resulting index remains well-defined in $[0,1]$ - that we do not pursue here. The Houtman--Maks index, by contrast, is geometry-free: it counts the largest fraction of rounds jointly rationalizable, with no reference to a deflation point, and is derived directly from Theorem~\ref{thm: peak oriented}.

\subsection{The Consistency Index}

Following \cite{houtman_maks}, we define the consistency index as the largest fraction of rounds that can be jointly rationalized.

\begin{definition}[Consistency index]\label{def: consistency index}
For a dataset $D = \{\mathbf{q}^r, \mathcal{B}^r\}_{r \in \mathcal{R}}$, the \emph{consistency index} is
\begin{equation}\label{eq: consistency index}
\kappa^*(D) := \max_{T \subseteq \mathcal{R}} \frac{|T|}{R} \quad \text{subject to the system~\eqref{eq: afriat1}--\eqref{eq:po_nondegen} being feasible on } T.
\end{equation}
\end{definition}

The consistency index answers: what is the largest fraction of rounds that can be jointly rationalized by a peak-oriented concave utility with a common ideal point? When $\kappa^* = 1$, all rounds are jointly consistent. When $\kappa^* < 1$, some rounds must be removed to recover a common ideal-point structure. The maximization is over the subset $T$; the peak $\mathbf{y}^*$ and the Afriat numbers $(U^0, U^r, \lambda^r)$ are existentially quantified inside the feasibility constraint.

Computationally, we approximate $\kappa^*(D)$ by $\kappa^*_{h,\varepsilon,\mathcal{M}}(D)$: the peak space is discretized on a grid $G_h \subset X$, and the strict directional condition $\mathbf{g}^r \cdot (\mathbf{y} - \mathbf{q}^r) > 0$ is normalized to an $\varepsilon$-threshold. Under the coordinatewise orientation, this normalized condition takes the form $\sum_{s:\,y_s \ne q^r_s} \operatorname{sign}(y_s - q^r_s) g^r_s \ge \varepsilon$ (Appendix~\ref{appendix: consistency computation}). Following the mixed-integer linear programming approach to Houtman--Maks-type indices introduced by \citet{HEUFER201587} and extended by \citet{Demuynck2023} to a broader family of goodness-of-fit measures, our MILP uses binary indicators $\delta_r \in \{0,1\}$ for whether round $r$ is retained, continuous variables $(U^0, U^r, \lambda^r)$, and big-$\mathcal{M}$ linearization of the Afriat inequalities; the peak-oriented sign condition and directional nondegeneracy add constraints specific to our setting. For $S = 2$ with $R = 19$ rounds, each MILP has $19$ binary variables and is solved in well under a second per subject using Gurobi.

\subsection{The Nonparametric Peak Set}\label{subsection: peak set}

The consistency index $\kappa^*(D)$ records the maximal fraction of rounds that can be jointly rationalized at some peak; it is silent on where that peak lies. This subsection uses the fixed-peak reduction of Section~\ref{subsection: geometric content} (Proposition~\ref{prop: reduction}) to define and characterize the set of ideal points the data can support.

\paragraph{Fixed-peak consistency.}
For $\mathbf{y}\in X$, let
\[
  \kappa(\mathbf{y};D):=\max\Bigl\{\tfrac{|T|}{R}:\ T\subseteq\mathcal{R}
  \text{ is peak-oriented rationalizable at }\mathbf{y}\Bigr\}
\]
be the largest fraction of rounds rationalizable when the peak is held fixed at $\mathbf{y}$ (in the sense of Section~\ref{subsection: geometric content}). The consistency index of Definition~\ref{def: consistency index} then decomposes as
\begin{equation}\label{eq: kappa decomposition}
  \kappa^*(D)=\max_{\mathbf{y}\in X}\kappa(\mathbf{y};D),
\end{equation}
since optimizing jointly over $(T,\mathbf{y})$ equals optimizing over $\mathbf{y}$ the best fraction attainable at $\mathbf{y}$.\footnote{The maximum is attained even though $\kappa(\cdot;D)$ is not continuous: it takes values in the finite set $\{0, 1/R, \ldots, 1\}$, so its supremum over $X$ lies in that set and is achieved by some $\mathbf{y} \in X$. In particular the Houtman--Maks peak set $\mathcal{Y}^*(D)$ defined below is nonempty.} The decomposition makes explicit a two-step structure: for each candidate ideal point we ask how much of the data it rationalizes, and only then optimize over the ideal point.

\paragraph{The nonparametric peak set.}
For $\alpha\in[0,1]$, the level set of admissible peaks is
\[
  \mathcal{Y}_\alpha(D):=\{\mathbf{y}\in X:\kappa(\mathbf{y};D)\ge\alpha\}
  =\bigcup_{\substack{T\subseteq\mathcal{R}\\ |T|/R\ge\alpha}}\mathcal{Y}_T(D),
\]
where $\mathcal{Y}_T(D)$ is the set of peaks at which $T$ is peak-oriented rationalizable. Two cases matter: the \emph{exact identified set} $\mathcal{Y}_1(D)$ (possibly empty), and the \emph{Houtman--Maks peak set} $\mathcal{Y}^*(D):=\{\mathbf{y}:\kappa(\mathbf{y};D)=\kappa^*(D)\}$. We emphasize that $\mathcal{Y}^*(D)$ is not an identified set for a globally single-peaked utility: it is the set of candidate peaks satisfying the finite-data peak-oriented implications of concave ideal-point rationalization. Since any single-peaked rationalization satisfies these implications (Proposition~\ref{prop: sp implies po}), $\mathcal{Y}^*(D)$ contains every peak that would survive a stronger global single-peakedness test, but does not pin down a unique ideal answer.

The peak set inherits the geometry of the double cones in Lemma~\ref{lem: double cone}: each round contributes a union of two opposite orthants, and, away from the finite set of observed choices, the level set $\mathcal{Y}_\alpha(D)$ is a finite union of orthant intersections on which the oriented data satisfy the signed-normal GARP cycle condition. Candidate peaks that coincide with an observed choice are checked directly by the Afriat system or MILP. As a consequence, the peak set may fail to be convex or connected.\footnote{Proposition~\ref{prop: cells} in Appendix~\hyperref[appendix: proofs]{A} gives the cone--GARP cell decomposition away from the finite set of observed choices.}

\begin{remark}[Centroid]\label{rem: centroid}
The centroid of the grid peak set $\mathcal{Y}^*_h(D)$ is a descriptive summary of where the maximally rationalizing peaks lie. Since that set may be nonconvex or disconnected, the centroid need not itself be admissible and should not be read as a nonparametric point estimate of the peak. The corner-anchored budgets ``triangulate'' the candidate peak through the intersection of round-by-round double cones without singling out a unique ideal answer.
\end{remark}

\section{Application}\label{section: application}

\subsection{Data Description}\label{subsection: data}

We implemented a PSM to study political preferences. The data were collected through the Qualtrics platform in the week following the 2024 European Parliament election in France (June 19--30, 2024), at a high-salience political moment: the Rassemblement National had won a large share of votes, about $31\%$, and the French President dissolved the National Assembly. The two questions in the PSM map onto the cleavage that election dramatized -- support for disadvantaged groups versus support for groups that contribute economically - so the application is a substantive setting for the methodology as well as a technical illustration. Of the 476 respondents who completed the PSM portion of the survey, we exclude 3 with reported monthly income above $\text{\texteuro}15{,}000$ (likely respondents who entered annual figures where monthly was asked) and 4 who left the occupation question blank, leaving an analytical sample of $N = 469$ respondents (Table~\ref{tab:summary stats}).

The experiment is not monetarily incentivized. There is no objectively ``correct'' survey answer to reward, and the natural alternative - paying for consistency itself - would reward the very behavior the consistency index is meant to measure. It might also push respondents toward the single-peaked structure the index is designed to test. We therefore do not incentivize, in keeping with standard practice in the survey literature \citep{Stantcheva2022}. The cost is that low consistency cannot be cleanly attributed to genuinely non-single-peaked preferences as opposed to low effort or confusion with the interface. The consistency and power indices quantify and flag this - uniformly random answers are detected as inconsistent for every subject's budget configuration (the power index equals 1.00; Section~\ref{subsection: rationality results}) - but they cannot by themselves separate the two channels.

\subsection{Experimental Setup}\label{subsection: setup}

Respondents answered two questions measuring policy priorities:

\begin{tcolorbox}[colback=blue!5!white, colframe=black!75!black, title=PSM: Policy Priorities]
\begin{itemize}
\item Question 1: ``On a scale from 0 to 10, to what extent do you agree: `Policies should support disadvantaged groups in society.'\,''
\item Question 2: ``On a scale from 0 to 10, to what extent do you agree: `Policies should support the groups that contribute to the economy.'\,''
\end{itemize}
\end{tcolorbox}

The implementation proceeds in two stages. In an initial unconstrained round (round~0), each respondent answers the two questions freely, with no constraint on the choice set. This stated answer $\mathbf{q}^0$ serves two purposes: it calibrates the budget levels in subsequent rounds, and it provides a benchmark for comparison with the revealed ideal answer. In the 19 constrained rounds that follow ($R = 19$), respondents face linear budget constraints with varying corners and prices. The order of the constrained rounds was randomized at the participant level, and the order of the two questions was randomized across participants. At the start of each constrained round, a preliminary answer was randomly drawn from the choice set, and participants adjusted it using sliders. The slider movements were co-dependent, ensuring that the displayed answer always satisfies the budget constraint. Each corner of $[0,10]^2$ is associated with four or five rounds. Table~\ref{table: design} gives the origins and price vectors for each round.

The budget level in round $r$ is calibrated to the respondent's stated answer:
\begin{equation}
    m^r = \max(\mathbf{q^0_{o^r}\cdot p^r}-4,\ 3),
\end{equation}
where the offset of 4 scale points positions the budget line at a substantial distance from $\mathbf{q}^0$ (40\% of the $[0,10]$ scale), and the $\max(\cdot, 3)$ ensures the budget line remains within $[0,10]^2$. This calibration ensures that the budget lines generate informative trade-offs for each respondent, since they are centered around the region of the answer space where the respondent's preferences are most relevant.

\subsubsection*{Design considerations}

A distinctive feature of this implementation is that budget levels are endogenous. They depend on the respondent's own stated answer, unlike in standard revealed-preference experiments \citep{andreoni2002, choi2014_rationality, fisman2015_science}, where budgets are assigned exogenously by the experimenter. Two consequences follow. First, two respondents with different stated answers face different budget sets, so cross-individual comparisons of rationality scores and utility parameters are based on different ``experiments.'' Second, if the stated answer is affected by cognitive biases, those biases propagate into the budget levels.

As established in Section~\ref{section: theory}, the characterization results hold for any collection of linear budget sets, regardless of how budget levels are determined. A candidate peak is recovered as part of the feasibility system from the constrained choices alone: the pattern of choices across budget lines with different corners and prices disciplines the location of the candidate peak. The peak-oriented sign condition substantially constrains the peak location, but does not deliver a point estimate. The consistency index measures how well the data conform to peak-oriented rationalization; the parametric estimation (Section~\ref{subsection: estimation}) produces a point-valued estimate under the quadratic functional-form restriction. An alternative design could set budget levels exogenously, at the cost of losing personalization: respondents with extreme preferences would face less informative trade-offs. The current implementation trades exogeneity for efficiency, using the stated answer as a calibration device. The stated answer is not used in the Afriat inequalities. It only determines the budget levels. Its dual role as both a calibration tool and a ``stated ideal answer'' for comparison with the revealed ideal should be kept in mind when interpreting the results.

\subsection{Rationality Assessment}\label{subsection: rationality results}

Following \citet{bronars}, we assess the discriminatory power of the PSM design by simulating random behavior under each subject's budget configuration. We generate $100$ synthetic datasets by drawing uniformly random choices on each budget line $\mathcal{B}^r$, and define the \emph{power index} as the fraction of synthetic datasets for which the consistency index is strictly below~1. A high power index indicates that random behavior is reliably detected as inconsistent.\footnote{\citet{cherchye_2025} develop a more rigorous permutation-based statistical test of utility maximization, with finite-sample Type-I error control and asymptotic power one against approximate utility maximization. We report the simpler Bronars-style power index here, which suffices to confirm that uniform random choices on each budget line never achieve perfect rationalization in our design ($\kappa^* < 1$ for every synthetic dataset). Adapting the permutation framework of \citet{cherchye_2025} to the corner-anchored, peak-oriented Afriat system would deliver per-subject $p$-values; we leave this extension for future work. See also \citet{beatty2011} for an alternative diagnostic in the revealed-preference literature.}

Approximately $35\%$ of constrained-round observations have a coordinate at $0$ or $10$ --- a respondent's slider resting at an endpoint of the budget segment when that endpoint lies on the boundary of $X$. Under the standing budget-interiority assumption, Lemma~\ref{lem:normal-supergradient} guarantees the existence of a supergradient parallel to $\mathbf{a}^r$ at every such choice, so Theorem~\ref{thm: peak oriented} applies uniformly to interior and boundary observations.

Table~\ref{tab:sum stat rationality} reports summary statistics for four rationality diagnostics. Throughout this section, $\kappa^*$ refers to the computed object $\kappa^*_{h,\varepsilon,\mathcal{M}}(D)$ defined in Section~\ref{section: rationality} and Appendix \ref{appendix: consistency computation}, with grid step $h = 0.5$ and nondegeneracy tolerance $\varepsilon = 10^{-3}$. Using all 19 constrained rounds, the average consistency index is $\kappa^* = 0.82$: at the optimal peak, about 15 of 19 rounds can be jointly rationalized by a peak-oriented concave utility. The average power index is 1.00: uniformly random choices never achieve $\kappa^* = 1$, confirming that the PSM design reliably distinguishes optimizing from random behavior. The mean adjustment $\Delta D = 2.57$ scale points indicates that respondents move substantially away from the randomly drawn preliminary answer rather than submitting it as-is, so the observed choices reflect deliberate adjustment rather than acceptance of the default. The mean identified-set share of $0.015$ ($p_{50} = 0.007$) shows that the nonparametric peak set is highly concentrated - typically about $1\%$ of the candidate peak space - so the corner-anchored budgets localize the candidate peak to a small region for the median subject, even without parametric assumptions.

\subsection{Utility Estimation}\label{subsection: estimation}

\subsubsection*{From the nonparametric theorem to parametric estimation}

The consistency index records \emph{how much} of the data is rationalizable; the nonparametric peak set $\mathcal{Y}^*(D)$ (Section~\ref{subsection: peak set}) records \emph{where} the peak lies. Empirically we compute its grid analogue $\mathcal{Y}^*_h(D)$ on a grid $G_h$ (step size $h = 0.5$ on $X$, 441 points total).

The grid peak set is typically small and concentrated. The median subject has 3 grid points in $\mathcal{Y}^*_h(D)$, less than 1\% of the grid, and a median range of 0.5 scale points in each dimension --- a single grid step. The corner-anchored budgets thus triangulate the peak to a small region even without parametric assumptions (Lemma~\ref{lem: double cone}). The peak set need not be connected: 35\% of subjects (162 of 469) have a disconnected $\mathcal{Y}^*_h(D)$, where separate regions achieve the same maximal $\kappa^*$ through different retained subsets.\footnote{Halving the grid step to $h=0.25$ changes $\kappa^*$ for 6\% of subjects (28 of 469), with a mean increase of $0.003$, confirming that the $h=0.5$ resolution is adequate.}

Because $\mathcal{Y}^*_h(D)$ may be disconnected, its centroid $\hat{\mathbf{y}}^*$ (the mean of the grid points in $\mathcal{Y}^*_h(D)$) need not belong to the set; it is a descriptive summary of where the nonparametric system places the peak, not a point estimate.

\subsubsection*{Parametric estimation}

We fit the following single-peaked utility to the data:
\begin{equation}\label{eq: utility smooth}
    u^i(q_1, q_2) = -\frac{1}{2} a_1^i \left( q_1 - b_1^i \right)^{2} -\frac{1}{2} a_2^i \left( q_2 - b_2^i \right)^{2},
\end{equation}
where $b_k^i \in [0, M_k]$ is the ideal answer to question $k$ (the peak of $u^i$ on $X$) and $a_k^i > 0$ measures the importance of question $k$ (strict positivity is required for a unique peak at $(b_1^i, b_2^i)$).\footnote{The specification is a monotone transformation of the negative weighted $L_\rho$-norm $-(a_1^i |q_1 - b_1^i|^{\rho} + a_2^i |q_2 - b_2^i|^{\rho})^{1/\rho}$ with $\rho = 2$ (ordinally equivalent, not the same cardinal formula). The general-$\rho$ form is concave for $\rho \ge 1$ and fails to be concave generally for $\rho < 1$.} With separable utility, indifference curves are axis-aligned ellipses centered on $(b_1^i, b_2^i)$.

Beyond identification, the elliptical specification carries an informal interpretation of how internal states map to recorded answers. The importance parameter $a_k^i$ encodes that questions may differ in centrality to the respondent: a higher $a_k^i$ means a steeper utility cost of deviating from the ideal answer $b_k^i$ in dimension $k$. The quadratic shape implies that this cost grows with the squared deviation, so the marginal cost of distorting an answer increases linearly in the deviation from $b_k^i$. These are propositions about the substantive content of the functional form, not derivations from a deeper psychological theory of survey response; they describe what the specification asserts about the cost of compromising on different questions when a budget binds.

The predicted answer in round $r$, when the unconstrained first-order solution lies in the relative interior of the budget segment $\mathcal{B}^r \cap X$, is:
\begin{equation}\label{eq: marshallian demand}
    \hat{q}^r_{1, o^r}=   \alpha^r b_{1,o^r}+(1-\alpha^r)\frac{R^r-p_2^r b_{2,o^r}}{p_1^r}, \quad \hat{q}^r_{2, o^r}= \frac{R^r-p^r_1 \hat{q}^r_{1, o^r}}{p^r_2},
\end{equation}
where $\alpha^r= \frac{a_1/(p_1^r)^2}{a_1/(p_1^r)^2+a_2/(p_2^r)^2}$ and $R^r = m^r$ is the budget in round $r$. In corner coordinates anchored at $o^r$, the first-order condition equates the marginal rate of substitution to the price ratio: $\frac{a_1(q_{1,o^r} - b_{1,o^r})}{a_2(q_{2,o^r} - b_{2,o^r})} = \frac{p_1^r}{p_2^r}$. When the first-order solution lies outside the segment, the constrained maximizer is the endpoint of $\mathcal{B}^r \cap X$ with the highest utility, and the estimation projects the unconstrained solution onto $\mathcal{B}^r \cap X$.

\paragraph{Identification and normalization of the importance parameters.}
Inspection of~\eqref{eq: marshallian demand} shows that $\alpha^r$ depends on $(a_1, a_2)$ only through the ratio $a_1/a_2$: rescaling both importance parameters by any positive constant leaves every predicted choice, and therefore the NLS objective, unchanged. The levels of $a_1$ and $a_2$ are not identified from the data, only their relative weight. We therefore impose the normalization
\[
a_1^i + a_2^i = 1, \qquad a_1^i, a_2^i \in (0, 1),
\]
under which $a_1^i$ is the share of importance respondent $i$ places on Q1 and $a_2^i = 1 - a_1^i$ is the corresponding share for Q2. The parameters estimated per respondent (after normalization) are $(a_1^i, b_1^i, b_2^i)$, with $a_2^i = 1 - a_1^i$ implied. The normalization removes only the common-scale indeterminacy in $(a_1, a_2)$; we do not claim global point identification. Because $a_1^i$ now lives on a common scale across respondents, it supports interpersonal comparison of importance, which the unnormalized version did not.

We estimate $(a_1^i, b_1^i, b_2^i)$ for each individual $i$ by minimizing the sum of squared prediction errors across the 19 constrained rounds:
\begin{equation}
    \min_{a_1,\, b_1,\, b_2} \sum_{r=1}^{R} \left( q^r_{1, o^r} - \hat{q}^r_{1, o^r} \right)^2 + \left( q^r_{2, o^r} - \hat{q}^r_{2, o^r} \right)^2,
\end{equation}
using a nonlinear least squares algorithm with constraints $a_1 \in [\eta, 1-\eta]$, $a_2 = 1 - a_1$, and $b_1, b_2 \in [0, 10]$, where $\eta = 10^{-4}$. The $b$-bounds enforce that the ideal answer lies in the answer space, consistent with the Section~\ref{section: theory} convention that the peak $\mathbf{y}^* \in X$; this is also an identification condition, since outside $X$ many $b$ values map to the same constrained maximizer at a corner of $\mathcal{B}^r \cap X$. The normalization removes the common-scale indeterminacy and yields normalized NLS estimates of $(a_1^i, b_1^i, b_2^i)$ for every respondent. We do not establish global point identification analytically. In multistart numerical checks on the realized design, all starting values converged to the same reported solution within the stated tolerance for every respondent, which provides evidence of numerical stability rather than a formal identification result. Tables~\ref{tab:result PSM q1}--\ref{tab:result PSM q2} report on the full sample.

\subsection{Results}\label{subsection: results}

\subsubsection*{Stated answer, nonparametric peak-set centroid, and parametric ideal}

\begin{figure}[!htbp]
    \centering
    \includegraphics[width=\linewidth]{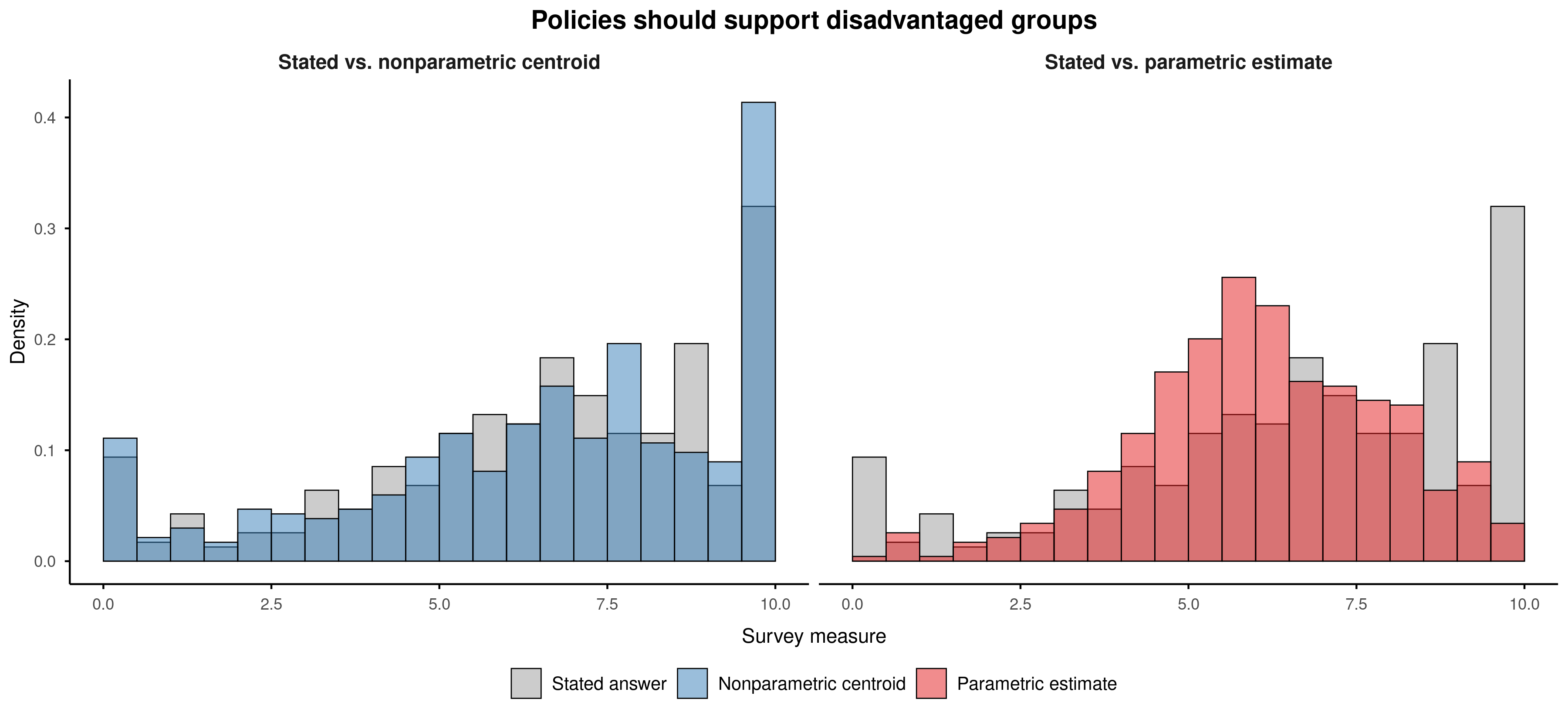}
    \caption{Distribution of the stated answer $q_1^0$ (grey), the nonparametric centroid $\hat{y}_1^*$ (blue, left panel), and the parametric estimate $b_1$ (red, right panel). All three measures are shown as histograms with binwidth $0.5$; the stated answer appears in both panels as a common baseline. The nonparametric centroid is the mean of the grid points in the Houtman--Maks peak set $\mathcal{Y}^*_h(D)$.}
    \label{fig:density: minority}
\end{figure}

\begin{figure}[!htbp]
    \centering
    \includegraphics[width=\linewidth]{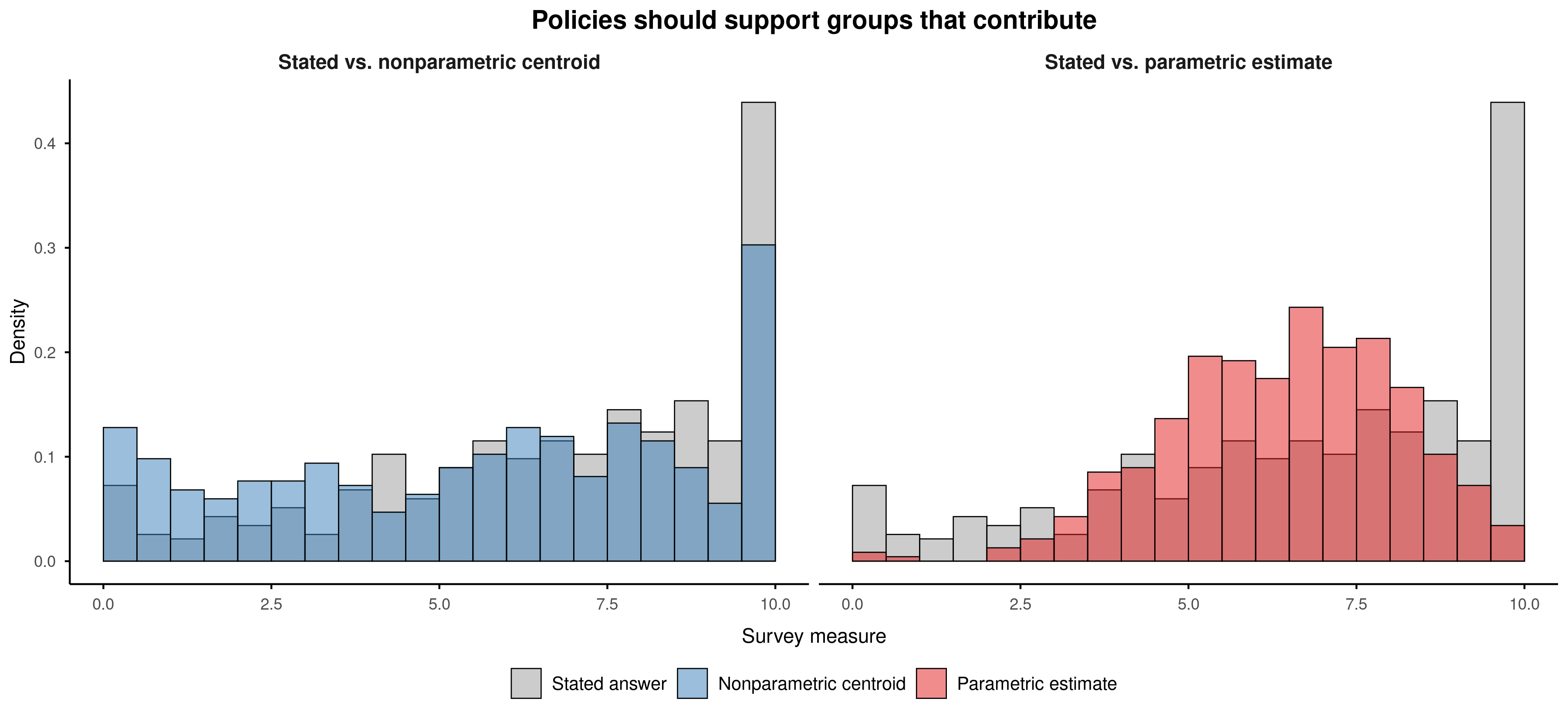}
    \caption{Distribution of the stated answer $q_2^0$ (grey), the nonparametric centroid $\hat{y}_2^*$ (blue, left panel), and the parametric estimate $b_2$ (red, right panel). All three measures are shown as histograms with binwidth $0.5$; the stated answer appears in both panels as a common baseline.}
    \label{fig:density contributors}
\end{figure}

Figures~\ref{fig:density: minority} and~\ref{fig:density contributors} compare three summary measures of where each respondent's ideal answer lies: the stated answer $q_k^0$, the nonparametric peak-set centroid $\hat{y}_k^*$ (mean of the grid points in $\mathcal{Y}^*_h(D)$ along dimension $k$; a descriptive summary, not a point estimate; see Remark~\ref{rem: centroid}), and the parametric ideal-answer estimate $b_k$ (the only one of the three that is a model-based point estimate of the ideal answer). All three lie in $[0,10]$ in the data, with the parametric estimate constrained to the answer space by the bounds in the NLS estimation (Section~\ref{subsection: estimation}).

The three distributions place  different mass at the scale endpoints. The stated answer shows pronounced clustering at $10$: $14\%$ of respondents report $q_1^0=10$ and $17\%$ report $q_2^0=10$. The nonparametric centroid is more spread out but still places noticeable mass at the endpoint. $19\%$ of subjects have $\hat{y}_1^*=10$ and $14\%$ have $\hat{y}_2^*=10$, reflecting respondents whose Houtman--Maks peak set genuinely reaches the corner of the answer space. The parametric estimate $b_k$, by contrast, is smoothly distributed in the interior of $[0,10]$ with essentially no mass at the endpoints: the closest $b_1$ estimate to a boundary is $0.037$ below $10$, and no $b_1$ estimate falls within $0.03$ of either $0$ or $10$.

This progression - stated answer, then nonparametric centroid, then parametric estimate - shows the cumulative effect of the rationality assumption and the functional form. That the nonparametric centroid tracks the stated answer more closely than the parametric estimate does may partly be a consequence of the calibration design (Section~\ref{subsection: setup}): budget lines are positioned a fixed distance from $\mathbf{q}^0$, so the observed choices, and the double cones emanating from them, are anchored near the stated answer. The parametric model then selects a point estimate under the quadratic functional-form assumption and recovers the paired importance weights $a_k$ that the nonparametric system is silent about. The elliptical form has a specific reason to prefer interior peaks over boundary peaks. With $b_k$ pinned at $10$, the constrained maximizer on every budget segment essentially clamps to the segment's high-$k$ endpoint, so predicted answers vary very little across rounds; with an interior $b_k < 10$ combined with a large importance weight $a_k$, the same pull toward high-$k$ answers is achieved while the first-order condition~\eqref{eq: marshallian demand} still generates round-to-round variation as prices and budgets change. When observed answers show any variation across rounds - as they do for almost every respondent - least squares prefers the interior-peak specification because it can fit that variation, while the boundary-peak specification cannot.

\subsubsection*{Sociodemographic determinants}

Tables~\ref{tab:result PSM q1} and~\ref{tab:result PSM q2} report the sociodemographic determinants of all four survey measures. The stated answer $q_k^0$ (col.~1), the nonparametric centroid $\hat{y}_k^*$ (col.~2), the importance share $a_k$ (col.~3), and the parametric ideal answer $b_k$ (col.~4). Under the normalization $a_1 + a_2 = 1$, the col.~3 coefficients for Q1 and Q2 are mirror images of each other. Throughout this subsection, ``Q1, col.~$j$'' refers to column~$j$ of Table~\ref{tab:result PSM q1} and ``Q2, col.~$j$'' to column~$j$ of Table~\ref{tab:result PSM q2}.

\emph{Order effects.} The order in which questions are presented significantly affects stated answers: $q_2^0$ decreases by about 0.5 scale points when question~1 is presented first (Q2, col.~1, $p < 0.10$) and the importance share $a_1$ rises by about 0.08 (Q1, col.~3, $p < 0.01$, equivalently $a_2$ falls by the same amount). The nonparametric peak in dimension 1 and the parametric ideal answer in dimension 2 also show marginally significant order effects ($\hat{y}_1^* = -0.46$, Q1 col.~2, $p < 0.10$; $b_2 = -0.28$, Q2 col.~4, $p < 0.10$), while the corresponding orthogonal-dimension coefficients ($\hat{y}_2^*$, $b_1$) are not significant at conventional levels. The order effect therefore operates primarily through the importance channel and only weakly through the recovered ideal answer, consistent with the interpretation that $q^0$ conflates stance and importance.

\emph{Political affiliation.} Extreme Right voters state about 0.9 scale points less support for disadvantaged groups (Q1, col.~1, $p < 0.05$). The nonparametric peak picks up a distinct pattern: Extreme Right voters have a significantly lower $\hat{y}_2^*$ ($-1.0$, Q2, col.~2, $p < 0.05$). Their parametric ideal answer $b_1$ is also significantly lower ($-0.73$, Q1, col.~4, $p < 0.01$). Extreme Left voters state about 2 points more support for disadvantaged groups (Q1, col.~1, $p < 0.01$), and their parametric ideal answer to supporting contributors is about 1.2 points lower (Q2, col.~4, $p < 0.01$). Left voters assign a significantly higher importance share to question~1 ($+0.08$, Q1, col.~3, $p < 0.05$).

\emph{Gender and education.} Women agree less with supporting economically contributing groups in stated answers ($-0.58$, Q2, col.~1, $p < 0.05$). Highly educated respondents have significantly higher parametric ideal answers for both questions (col.~4 of both tables, $p < 0.05$) and higher nonparametric peaks (col.~2 of both tables, $p < 0.05$), patterns largely absent from their stated answers (col.~1), suggesting that the PSM captures preference heterogeneity masked in standard survey measures.

\subsubsection*{Interpersonal comparison via importance shares}

The importance share $a_1^i$ (and equivalently $a_2^i = 1 - a_1^i$) provides a basis for comparing how respondents weight Q1 relative to Q2. This information is unavailable in standard surveys. For example, Left voters assign a higher share to question~1 (Q1, col.~3, $p < 0.05$; equivalently a lower share to question~2, Q2 col.~3) while their ideal answers do not differ significantly from centrist voters. Presenting question~1 first also shifts the share toward Q1 by about 8 percentage points (Q1 col.~3, $p < 0.01$). These patterns suggest that political affiliation and presentation order shape which question respondents prioritize, not only what they answer.

\section{Conclusion}\label{section: discussion}

This paper develops an Afriat-type theorem for concave rationalization with an unknown ideal point under corner-anchored budgets, and applies it to survey methodology. The main theoretical contribution is a system of Afriat inequalities where the unknown peak enters as a virtual observation with the highest utility. The peak-oriented version of the theorem adds the requirement that supergradients at observed choices point coordinatewise toward the peak, providing a testable necessary condition for single-peaked rationalizability. Two empirical objects emerge from this characterization. The consistency index records the largest fraction of rounds that can be jointly rationalized by a peak-oriented concave utility with a common ideal point. The nonparametric peak set collects the candidate ideal answers the data can support without any parametric assumption.

The Priced Survey Methodology (PSM) brings revealed-preference tools to bear on survey responses by translating the assumption of a single ideal answer into a finite, testable system on constrained choices. The nonparametric peak set substantially constrains the candidate ideal point as part of the feasibility system, though it does not deliver a unique peak. A parametric single-peaked model sharpens this into point estimates of ideal answers and importance shares. The parametric ideal answer is continuous-valued on the answer space, recovers the importance shares that the nonparametric system is silent about, and - in the application studied here - displays substantially weaker order effects than the stated answers, though one coordinate retains a marginally significant order coefficient.

Several directions for future research are worth noting. First, while this paper implements a PSM with two questions, the theory extends to any number of questions. Scaling the PSM to higher-dimensional surveys - and designing practical interfaces that elicit informative trade-offs over many questions at once - is an open challenge. Second, exploring alternative measures of approximate rationality such as statistical tests for noise \citep{aguiar2020} or approximate utility maximization \citep{dziewulski2021} could complement the Houtman--Maks consistency index and further improve preference recovery from PSM data.

\bibliographystyle{apalike}
\bibliography{bibliography}

\clearpage

\clearpage
\section*{Tables}\label{section: tables and figures}

\begin{table}[!htbp]
\centering
\caption{Round design: origin and price assignments across the 19 constrained rounds}\label{table: design}
\begin{minipage}{0.45\textwidth}
\centering
{\small (a) Origin assignment}\\[2pt]
\begin{tabular}{cc}
\toprule
Round $r$ & Origin $\mathbf{o^r}$\\
\midrule
1--5 & (0,0) \\
6--10 & (10,0) \\
11--15 & (0,10) \\
16--19 & (10,10) \\
\bottomrule
\end{tabular}
\end{minipage}
\hfill
\begin{minipage}{0.45\textwidth}
\centering
{\small (b) Price assignment}\\[2pt]
\begin{tabular}{cc}
\toprule
Round $r$ & Prices $\mathbf{p^r}$ \\
\midrule
1, 6, 11, 16 & (2, 1) \\
2, 7, 12, 17 & (1, 2) \\
3, 8, 13, 18 & ($\frac{2}{3}$, 3) \\
4, 9, 14, 19 & (3, $\frac{2}{3}$) \\
5, 10, 15 & ($\frac{5}{4}$, $\frac{5}{4}$) \\
\bottomrule
\end{tabular}
\end{minipage}
\end{table}

\begin{table}[!htbp]
\centering
\begin{threeparttable}
\centering
\begin{tabular}{lc}
\hline \hline
Variables & Number of Participants \\
\hline
\emph{Female} & 241 \\
\addlinespace
\emph{Age} & \\
18-34 & 129 \\
35-49 & 136 \\
50-64 & 151 \\
65+ & 53 \\
\addlinespace
\emph{Education} & \\
Low & 140 \\
Medium & 246 \\
High & 83 \\
\addlinespace
\emph{Monthly income} & \\
\texteuro0-2500 & 202 \\
\texteuro2500-3499 & 91 \\
\texteuro3500-4999 & 113 \\
\texteuro5000+ & 63 \\
\addlinespace
\emph{Media Exposure (Hours/day)} & \\
<1 & 42 \\
1-3 & 179 \\
3-6 & 145 \\
6-9 & 60 \\
9+ & 43 \\
\addlinespace
Observations & 469 \\
\hline
\end{tabular}
\begin{tablenotes}
\small
\item \emph{Notes.} Education is recorded as the respondent's highest completed qualification and grouped into three categories: \emph{Low} (no diploma, primary education, lower-secondary education, or vocational baccalaureate); \emph{Medium} (general or technological baccalaureate, intermediate pre-tertiary diplomas, or short-cycle tertiary degrees at the Bac+2 or Bac+3 level, e.g.\ Bachelor, DUT, BTS); and \emph{High} (long-cycle tertiary degrees at the Bac+4 or Bac+5 level, e.g.\ Master, DEA, DESS, or Doctorat). Monthly income refers to self-reported net household income. Media exposure is the respondent's self-reported average daily time spent consuming news media.
\end{tablenotes}
\end{threeparttable}

\caption{Sociodemographic Variables.}
\label{tab:summary stats}
\end{table}

\begin{table}[!htbp]
\centering
\begin{threeparttable}
\centering
\begin{tabular}{lcccc}
\hline \hline
 & $\Delta D$ & $\kappa^*$ & Power & Id.\ set \\
 & (1) & (2) & (3) & (4) \\
\hline
Mean & 2.57 & 0.82 & 1.00 & 0.015 \\
Std & 1.35 & 0.09 & 0.00 & 0.020 \\
p5 & 0.48 & 0.68 & 1.00 & 0.002 \\
p25 & 1.30 & 0.74 & 1.00 & 0.005 \\
p50 & 2.73 & 0.79 & 1.00 & 0.009 \\
p75 & 3.63 & 0.89 & 1.00 & 0.018 \\
p95 & 4.67 & 1.00 & 1.00 & 0.051 \\
\hline
Fraction $\kappa^* = 1$ & & 0.06 & & \\
\hline
\end{tabular}
\begin{tablenotes}
\small
\item $\Delta D$: mean Euclidean distance, across rounds, between the randomly drawn preliminary answer and the respondent's submitted answer; large values indicate substantial adjustment away from the default. $\kappa^*$: consistency index $\kappa^*_{h,\varepsilon}(D)$ (Definition~\ref{def: consistency index}), the largest fraction of rounds jointly satisfying the peak-oriented Afriat system. Power: share of synthetic datasets, drawn by sampling uniformly random choices on each budget line under the respondent's budget configuration, for which $\kappa^* < 1$. Id.\ set: share of the grid $G_h$ achieving $\kappa^*$ (the nonparametric peak set as a fraction of the candidate peak space). Sample: the 469 respondents used in Tables 2, 4 and 5.
\end{tablenotes}
\end{threeparttable}

\caption{Summary Statistics: Rationality Scores}
\label{tab:sum stat rationality}
\end{table}

\begin{table}[!htbp]
\centering
\small
\renewcommand{\arraystretch}{0.85}
\begin{threeparttable}
\centering
\begin{tabular}{lcccc}
\hline\hline
\multicolumn{5}{c}{\emph{Question 1: Policies should support disadvantaged groups}} \\
\hline
 & $q_1^0$ & $\hat{y}_1^*$ & $a_1$ & $b_1$
 \\
 & (1) & (2) & (3) & (4)
 \\
\hline
 Order & $-$0.353 & $-$0.459$^{*}$ & 0.077$^{***}$ & $-$0.056
 \\
  & (0.251) & (0.269) & (0.024) & (0.169)
 \\
\addlinespace
 \emph{Vote (ref: Center)} & & & & \\
 \quad Right & $-$0.479 & $-$0.126 & $-$0.032 & $-$0.244
 \\
  & (0.379) & (0.406) & (0.036) & (0.256)
 \\
 \quad Extreme Right & $-$0.903$^{**}$ & $-$0.018 & 0.004 & $-$0.725$^{***}$
 \\
  & (0.414) & (0.443) & (0.039) & (0.279)
 \\
 \quad Extreme Left & 1.970$^{***}$ & $-$0.932 & 0.083 & 0.552
 \\
  & (0.712) & (0.763) & (0.067) & (0.481)
 \\
 \quad Left & 0.220 & $-$0.236 & 0.079$^{**}$ & 0.448$^{*}$
 \\
  & (0.364) & (0.389) & (0.034) & (0.245)
 \\
 \quad Did not vote & 0.269 & $-$0.292 & $-$0.009 & 0.061
 \\
  & (0.457) & (0.490) & (0.043) & (0.309)
 \\
\addlinespace
 \emph{Age (ref: 18--34)} & & & & \\
 \quad 35--49 & 0.288 & $-$0.038 & $-$0.034 & $-$0.239
 \\
  & (0.354) & (0.379) & (0.033) & (0.239)
 \\
 \quad 50--64 & 0.244 & 0.337 & $-$0.051 & $-$0.613$^{**}$
 \\
  & (0.381) & (0.407) & (0.036) & (0.257)
 \\
 \quad 65+ & $-$0.015 & 0.440 & $-$0.070 & $-$0.756$^{*}$
 \\
  & (0.641) & (0.686) & (0.060) & (0.432)
 \\
 Female & 0.257 & 0.013 & 0.002 & 0.092
 \\
  & (0.257) & (0.275) & (0.024) & (0.174)
 \\
\addlinespace
 \emph{Education (ref: Low)} & & & & \\
 \quad Medium & 0.535$^{*}$ & 0.866$^{***}$ & $-$0.005 & 0.319
 \\
  & (0.299) & (0.321) & (0.028) & (0.202)
 \\
 \quad High & 0.060 & 1.100$^{**}$ & $-$0.015 & 0.663$^{**}$
 \\
  & (0.405) & (0.434) & (0.038) & (0.273)
 \\
\addlinespace
 \emph{Activity (ref: Employed)} & & & & \\
 \quad Home Work & 0.145 & 1.149 & 0.023 & 0.823
 \\
  & (0.775) & (0.829) & (0.073) & (0.523)
 \\
 \quad Retired & $-$0.038 & $-$0.911$^{*}$ & 0.029 & 0.294
 \\
  & (0.475) & (0.508) & (0.045) & (0.321)
 \\
 \quad Student & 0.355 & 0.153 & 0.058 & 0.389
 \\
  & (0.567) & (0.608) & (0.053) & (0.383)
 \\
 \quad Other & 0.696 & 1.477$^{**}$ & $-$0.028 & $-$0.045
 \\
  & (0.613) & (0.657) & (0.058) & (0.414)
 \\
\hline
Observations & 469 & 469 & 469 & 469
 \\
$R^{2}$ & 0.078 & 0.062 & 0.079 & 0.103
 \\
Adjusted $R^{2}$ & 0.037 & 0.020 & 0.038 & 0.063
 \\
\hline\hline
\end{tabular}
\begin{tablenotes}
\small
\item Robust standard errors in parentheses. Significance: $^{*}\,p<0.10$, $^{**}\,p<0.05$, $^{***}\,p<0.01$. Controls included but not reported: log income, media exposure, number of other adults, and number of children. Sample: $N = 469$ respondents. The importance parameters are normalized by $a_1 + a_2 = 1$ (Section~\ref{subsection: estimation}); col.~3 reports the share for the question of interest, so the col.~3 coefficients in Table~\ref{tab:result PSM q1} and Table~\ref{tab:result PSM q2} are mirror images.
\end{tablenotes}
\end{threeparttable}

\caption{Sociodemographic determinants of survey measures, Question~1. Column~(1): stated answer $q_1^0$. Column~(2): nonparametric centroid $\hat{y}_1^*$ of the Houtman--Maks peak set. Column~(3): importance parameter $a_1$. Column~(4): parametric ideal answer $b_1$.}
\label{tab:result PSM q1}
\end{table}

\begin{table}[!htbp]
\centering
\small
\renewcommand{\arraystretch}{0.85}
\begin{threeparttable}
\centering
\begin{tabular}{lcccc}
\hline\hline
\multicolumn{5}{c}{\emph{Question 2: Policies should support groups that contribute economically}} \\
\hline
 & $q_2^0$ & $\hat{y}_2^*$ & $a_2$ & $b_2$
 \\
 & (1) & (2) & (3) & (4)
 \\
\hline
 Order & $-$0.485$^{*}$ & $-$0.313 & $-$0.077$^{***}$ & $-$0.280$^{*}$
 \\
  & (0.262) & (0.298) & (0.024) & (0.158)
 \\
\addlinespace
 \emph{Vote (ref: Center)} & & & & \\
 \quad Right & $-$0.086 & $-$0.418 & 0.032 & 0.363
 \\
  & (0.396) & (0.451) & (0.036) & (0.239)
 \\
 \quad Extreme Right & $-$0.049 & $-$1.027$^{**}$ & $-$0.004 & 0.064
 \\
  & (0.432) & (0.491) & (0.039) & (0.260)
 \\
 \quad Extreme Left & $-$1.152 & 0.766 & $-$0.083 & $-$1.221$^{***}$
 \\
  & (0.745) & (0.846) & (0.067) & (0.448)
 \\
 \quad Left & $-$0.872$^{**}$ & 0.201 & $-$0.079$^{**}$ & $-$0.294
 \\
  & (0.380) & (0.432) & (0.034) & (0.229)
 \\
 \quad Did not vote & $-$0.066 & $-$0.278 & 0.009 & $-$0.229
 \\
  & (0.478) & (0.543) & (0.043) & (0.288)
 \\
\addlinespace
 \emph{Age (ref: 18--34)} & & & & \\
 \quad 35--49 & 0.298 & $-$0.402 & 0.034 & 0.197
 \\
  & (0.370) & (0.420) & (0.033) & (0.223)
 \\
 \quad 50--64 & 0.593 & $-$0.511 & 0.051 & 0.343
 \\
  & (0.398) & (0.452) & (0.036) & (0.240)
 \\
 \quad 65+ & 0.697 & $-$1.307$^{*}$ & 0.070 & 0.479
 \\
  & (0.670) & (0.761) & (0.060) & (0.403)
 \\
 Female & $-$0.582$^{**}$ & 0.288 & $-$0.002 & $-$0.206
 \\
  & (0.269) & (0.306) & (0.024) & (0.162)
 \\
\addlinespace
 \emph{Education (ref: Low)} & & & & \\
 \quad Medium & $-$0.185 & 0.141 & 0.005 & 0.294
 \\
  & (0.313) & (0.356) & (0.028) & (0.188)
 \\
 \quad High & $-$0.031 & 1.239$^{**}$ & 0.015 & 0.592$^{**}$
 \\
  & (0.423) & (0.481) & (0.038) & (0.255)
 \\
\addlinespace
 \emph{Activity (ref: Employed)} & & & & \\
 \quad Home Work & 0.484 & 1.064 & $-$0.023 & $-$0.066
 \\
  & (0.810) & (0.920) & (0.073) & (0.488)
 \\
 \quad Retired & $-$0.116 & 1.179$^{**}$ & $-$0.029 & $-$0.268
 \\
  & (0.496) & (0.564) & (0.045) & (0.299)
 \\
 \quad Student & 0.523 & 0.470 & $-$0.058 & 0.097
 \\
  & (0.593) & (0.674) & (0.053) & (0.357)
 \\
 \quad Other & $-$0.118 & 0.702 & 0.028 & 0.775$^{**}$
 \\
  & (0.641) & (0.729) & (0.058) & (0.386)
 \\
\hline
Observations & 469 & 469 & 469 & 469
 \\
$R^{2}$ & 0.062 & 0.062 & 0.079 & 0.077
 \\
Adjusted $R^{2}$ & 0.020 & 0.020 & 0.038 & 0.036
 \\
\hline\hline
\end{tabular}
\begin{tablenotes}
\small
\item Robust standard errors in parentheses. Significance: $^{*}\,p<0.10$, $^{**}\,p<0.05$, $^{***}\,p<0.01$. Controls included but not reported: log income, media exposure, number of other adults, and number of children. Sample: $N = 469$ respondents. The importance parameters are normalized by $a_1 + a_2 = 1$ (Section~\ref{subsection: estimation}); col.~3 reports the share for the question of interest, so the col.~3 coefficients in Table~\ref{tab:result PSM q1} and Table~\ref{tab:result PSM q2} are mirror images.
\end{tablenotes}
\end{threeparttable}

\caption{Sociodemographic determinants of survey measures, Question~2. Column~(1): stated answer $q_2^0$. Column~(2): nonparametric centroid $\hat{y}_2^*$ of the Houtman--Maks peak set. Column~(3): importance parameter $a_2$. Column~(4): parametric ideal answer $b_2$.}
\label{tab:result PSM q2}
\end{table}

\pagebreak
\clearpage
\newpage
\section*{Figures}

\newpage

\bigskip \bigskip
\appendix
\onehalfspacing
\pagebreak
\newpage
\phantomsection
\makeatletter\def\@currentlabel{A}\makeatother
\label{appendix: proofs}
\begin{LARGE}
    \noindent\textbf{Appendix A: Proofs}
\end{LARGE}

\setcounter{table}{0}
\renewcommand{\thetable}{A.\arabic{table}}
\setcounter{figure}{0}
\renewcommand{\thefigure}{A.\arabic{figure}}
\setcounter{section}{0}
\renewcommand{\thesection}{A.\arabic{section}}
\setcounter{equation}{0}
\renewcommand*{\theequation}{A.\arabic{equation}}
\setcounter{theorem}{0}
\renewcommand{\thetheorem}{A.\arabic{theorem}}
\setcounter{proposition}{0}
\renewcommand{\theproposition}{A.\arabic{proposition}}
\setcounter{lemma}{0}
\renewcommand{\thelemma}{A.\arabic{lemma}}
\setcounter{definition}{0}
\renewcommand{\thedefinition}{A.\arabic{definition}}
\setcounter{remark}{0}
\renewcommand{\theremark}{A.\arabic{remark}}


\section{Preliminary: first-order condition at an observed budget optimum}

\begin{lemma}[FOC at an observed budget optimum]\label{lem:normal-supergradient}
Let $u: X \to \mathbb{R}$ be continuous concave. Suppose $\mathbf{q}^r \in X$ maximizes $u$ over $\mathcal{B}^r = \{\mathbf{x} \in X: \mathbf{a}^r \cdot \mathbf{x} = \mu^r\}$, with $\mathbf{a}^r \neq \mathbf{0}$. Then, under the standing nondegenerate-budget assumption, there exists $\lambda^r \in \mathbb{R}$ such that $\lambda^r \mathbf{a}^r \in \partial u(\mathbf{q}^r)$.
\end{lemma}

\begin{proof}
Define $F: \mathbb{R}^S \to (-\infty, +\infty]$ by $F(\mathbf{x}) = -u(\mathbf{x})$ for $\mathbf{x} \in X$ and $F(\mathbf{x}) = +\infty$ otherwise. Then $F$ is convex with effective domain $\mathrm{dom}\,F := \{\mathbf{x} \in \mathbb R^S : F(\mathbf{x}) < +\infty\} = X$, and unpacking definitions, $\partial F(\mathbf{q}^r) = -\partial u(\mathbf{q}^r)$ (the supergradient of $u$ at $\mathbf{q}^r$ is defined by the same inequality as the subgradient of $-u$ over $X$). Let $H^r := \{\mathbf{x} \in \mathbb{R}^S : \mathbf{a}^r \cdot \mathbf{x} = \mu^r\}$ denote the full hyperplane. For any set $C \subseteq \mathbb R^S$, let $I_C$ denote the \emph{convex-analysis indicator function} of $C$, given by $I_C(\mathbf x) = 0$ for $\mathbf x \in C$ and $I_C(\mathbf x) = +\infty$ otherwise; standard convex analysis gives $\partial I_C(\mathbf x) = N_C(\mathbf x)$, the normal cone to $C$ at $\mathbf x$. Since $F$ already carries the box constraint through its $+\infty$ extension outside $X$, we have $F + I_{H^r} = F + I_{\mathcal{B}^r}$ as functions on $\mathbb{R}^S$, and $\mathbf{q}^r$ minimizes this sum on $\mathbb{R}^S$, so Fermat's rule gives
\[
  \mathbf{0} \in \partial(F + I_{H^r})(\mathbf{q}^r).
\]
The standing nondegenerate-budget assumption of §\ref{subsection: choice sets} gives $\mathrm{ri}(\mathrm{dom}\,F) \cap \mathrm{ri}(\mathrm{dom}\,I_{H^r}) = \mathrm{int}(X) \cap H^r \neq \emptyset$, where $\mathrm{ri}(C)$ denotes the \emph{relative interior} of a convex set $C$ (its interior taken inside the smallest affine subspace containing $C$; for the full-dimensional box, $\mathrm{ri}(X) = \mathrm{int}(X)$; for the hyperplane, $\mathrm{ri}(H^r) = H^r$ since $H^r$ has empty ordinary interior in $\mathbb R^S$). The Moreau--Rockafellar sum rule \citep[Thm.~23.8]{rockafellar1970} then yields
\[
  \partial(F + I_{H^r})(\mathbf{q}^r) = \partial F(\mathbf{q}^r) + N_{H^r}(\mathbf{q}^r) = -\partial u(\mathbf{q}^r) + \mathrm{span}\{\mathbf{a}^r\},
\]
where the second equality uses $N_{H^r}(\mathbf{q}^r) = \mathrm{span}\{\mathbf{a}^r\}$: the normal cone to an affine hyperplane is the one-dimensional subspace spanned by its normal vector.
Hence $\mathbf{0} \in -\partial u(\mathbf{q}^r) + \mathrm{span}\{\mathbf{a}^r\}$, so there exists $\lambda^r \in \mathbb{R}$ with $\lambda^r \mathbf{a}^r \in \partial u(\mathbf{q}^r)$.
\end{proof}

\medskip
\begin{remark}[Constant utility and directional nondegeneracy]\label{rem: constant excluded}
Suppose $u \equiv \mathrm{const}$ and $\mathbf{g}^r = \lambda^r \mathbf{a}^r \in \partial u(\mathbf{q}^r)$. By budget nondegeneracy, choose $\mathbf{z} \in H^r \cap \mathrm{int}(X)$. Because $\mathbf{z}$ is interior, for all sufficiently small $t > 0$, $\mathbf{z} \pm t\mathbf{a}^r \in X$. The supergradient inequality for the constant utility gives
\[
0 \;\le\; \lambda^r \mathbf{a}^r \cdot \bigl(\mathbf{z} \pm t\mathbf{a}^r - \mathbf{q}^r\bigr).
\]
Since $\mathbf{z} \in H^r$ by construction and $\mathbf{q}^r \in H^r$ by observed feasibility, $\mathbf{a}^r \cdot (\mathbf{z} - \mathbf{q}^r) = 0$, so both
\[
0 \;\le\; \lambda^r\, t\, \|\mathbf{a}^r\|^2 \qquad \text{and} \qquad 0 \;\le\; -\lambda^r\, t\, \|\mathbf{a}^r\|^2
\]
must hold, forcing $\lambda^r = 0$ and hence $\mathbf{g}^r = \mathbf{0}$. Constant utility therefore fails Definition~\ref{def: peak oriented}(e) whenever at least one retained observation differs from the candidate peak.
\end{remark}


\section{Proof of Theorem~\ref{theorem: halevy}}\label{appendix proof theorem 1}
We prove $(i) \Rightarrow (ii) \Rightarrow (iii) \Rightarrow (i)$.

\medskip
\noindent\textbf{$(i) \Rightarrow (ii)$.}
Assume there exist $\mathbf{y}^* \in X$ and a continuous concave $u: X \to \mathbb{R}$ with $\mathbf{y}^* \in \arg\max_{\mathbf{x} \in X} u(\mathbf{x})$ and $\mathbf{q}^r \in \arg\max_{\mathbf{x} \in \mathcal{B}^r} u(\mathbf{x})$ for all $r$. Define $U^0 := u(\mathbf{y}^*)$ and $U^r := u(\mathbf{q}^r)$.

For each $l \in \mathcal{R}$, by Lemma~\ref{lem:normal-supergradient} there exists $\lambda^l \in \mathbb{R}$ with $\mathbf{g}^l := \lambda^l \mathbf{a}^l \in \partial u(\mathbf{q}^l)$. The supergradient inequality gives, for all $\mathbf{x} \in X$:
\begin{equation}\label{eq: supergradient}
    u(\mathbf{x}) \leq u(\mathbf{q}^l) + \mathbf{g}^l \cdot (\mathbf{x} - \mathbf{q}^l).
\end{equation}

Applying~\eqref{eq: supergradient} at $\mathbf{x} = \mathbf{q}^r$ gives~\eqref{eq: afriat1}. Since $\mathbf{y}^*$ maximizes $u$: $U^r \leq U^0$, giving~\eqref{eq: afriat2}. Applying~\eqref{eq: supergradient} at $\mathbf{x} = \mathbf{y}^*$ gives~\eqref{eq: afriat3}.

\medskip
\noindent\textbf{$(ii) \Rightarrow (iii)$.}
Assume~\eqref{eq: afriat1}--\eqref{eq: afriat3} hold with $\mathbf{g}^l = \lambda^l \mathbf{a}^l$ and define $u$ by~\eqref{eq: utility}. Since $u$ is the minimum of finitely many affine functions capped at $U^0$, it is continuous, concave, and piecewise affine.

\emph{$u(\mathbf{q}^r) = U^r$}: By~\eqref{eq: afriat1} and~\eqref{eq: afriat2}, $U^l + \mathbf{g}^l \cdot (\mathbf{q}^r - \mathbf{q}^l) \geq U^r$ for every $l$ and $U^0 \geq U^r$, hence $u(\mathbf{q}^r) \geq U^r$. Taking $l = r$: $U^r + \mathbf{g}^r \cdot \mathbf{0} = U^r$, so $u(\mathbf{q}^r) \leq U^r$.

\emph{$u(\mathbf{y}^*) = U^0 = \max_{\mathbf{x} \in X} u(\mathbf{x})$}: By~\eqref{eq: afriat3}, $U^l + \mathbf{g}^l \cdot (\mathbf{y}^* - \mathbf{q}^l) \geq U^0$ for every $l$, and the cap gives $u(\mathbf{y}^*) \leq U^0$; together $u(\mathbf{y}^*) = U^0$. The cap also gives $u(\mathbf{x}) \leq U^0$ for every $\mathbf{x} \in X$.

\emph{Rationalization}: For $\mathbf{x} \in \mathcal{B}^r$: $\mathbf{a}^r \cdot (\mathbf{x} - \mathbf{q}^r) = 0$, so the $r$-th term in~\eqref{eq: utility} evaluates to $U^r$. Therefore $u(\mathbf{x}) \leq U^r = u(\mathbf{q}^r)$.

\medskip
\noindent\textbf{$(iii) \Rightarrow (i)$.} The function $u$ constructed in (iii) is continuous concave with $\mathbf{y}^*$ as a maximizer in $X$ and $\mathbf{q}^r$ as a maximizer in $\mathcal{B}^r$ for every $r$, so $u$ provides a concave rationalization with ideal point.

\hfill $\blacksquare$


\section{Proof of Theorem~\ref{thm: peak oriented}}

The proof of Theorem~\ref{theorem: halevy} applies with two additions: the sign condition~\eqref{eq:po_afriat_sign} and the nondegeneracy condition~\eqref{eq:po_nondegen} propagate through each step.

\medskip
\noindent\textbf{$(i) \Rightarrow (ii)$.}
Definition~\ref{def: peak oriented}(c) supplies $\lambda^r \in \mathbb{R}$ with $\mathbf{g}^r := \lambda^r \mathbf{a}^r \in \partial u(\mathbf{q}^r)$; the supergradient inequality on $X$ gives~\eqref{eq: afriat1}--\eqref{eq: afriat3} exactly as in Theorem~\ref{theorem: halevy}. Definition~\ref{def: peak oriented}(d) gives $g^r_s(y^*_s - q^r_s) \geq 0$ for all $r, s$, which is~\eqref{eq:po_afriat_sign}. Definition~\ref{def: peak oriented}(e) gives~\eqref{eq:po_nondegen}.

\medskip
\noindent\textbf{$(ii) \Rightarrow (iii)$.}
Define $u$ by~\eqref{eq: utility} with $\mathbf{g}^l = \lambda^l \mathbf{a}^l$. The proofs that $u(\mathbf{q}^r) = U^r$, $u(\mathbf{y}^*) = U^0 = \max_{\mathbf{x} \in X} u(\mathbf{x})$, and rationalization are identical to Theorem~\ref{theorem: halevy}. For every $\mathbf{x} \in X$, $u(\mathbf{x}) \leq U^r + \mathbf{g}^r \cdot (\mathbf{x} - \mathbf{q}^r)$ (the $r$-th term in the min). With $u(\mathbf{q}^r) = U^r$, this gives $\mathbf{g}^r = \lambda^r \mathbf{a}^r \in \partial u(\mathbf{q}^r)$, verifying Definition~\ref{def: peak oriented}(c). The sign and nondegeneracy conditions carry over directly from~\eqref{eq:po_afriat_sign}--\eqref{eq:po_nondegen}.

\medskip
\noindent\textbf{$(iii) \Rightarrow (i)$.} Immediate from the properties listed.

\hfill $\blacksquare$


\section{Proof of Proposition~\ref{prop: sp implies po}}

Suppose $u$ is continuous concave and single-peaked (Definition~\ref{def: single peaked}) with unique peak $\mathbf{y}^*$ and rationalizes $D$. By Lemma~\ref{lem:normal-supergradient}, for each $r$ there exists $\lambda^r \in \mathbb{R}$ with $\mathbf{g}^r := \lambda^r \mathbf{a}^r \in \partial u(\mathbf{q}^r)$. We verify the sign and directional nondegeneracy conditions on $\mathbf{g}^r$ directly.

\medskip
\noindent\emph{Sign condition.} We show $g^r_s(y^*_s - q^r_s) \geq 0$ for all $r, s$. Fix $r$ and $s$.

\emph{Case $q^r_s < y^*_s$:} Since $q^r_s < y^*_s \le M_s$ and $q^r_s \ge 0$, the path $\mathbf{x}(t) = \mathbf{q}^r + t \mathbf{e}_s$ (with $\mathbf{e}_s$ the $s$-th standard basis vector) for $t \in [0, y^*_s - q^r_s]$ lies in $X$ (coordinate $s$ stays in $[0, M_s]$; other coordinates are unchanged). Choose corner $c$ by $c_j = 0$ if $q^r_j \leq y^*_j$, $c_j = M_j$ otherwise. Then $(\mathbf{q}^r)_c \leq \mathbf{x}(t)_c \leq (\mathbf{y}^*)_c$ componentwise. Single-peakedness gives $u(\mathbf{q}^r) \leq u(\mathbf{x}(t))$. The supergradient inequality $u(\mathbf{x}(t)) \leq u(\mathbf{q}^r) + \mathbf{g}^r \cdot (\mathbf{x}(t) - \mathbf{q}^r) = u(\mathbf{q}^r) + g^r_s t$ then gives $g^r_s t \ge 0$, so $g^r_s \geq 0$.

\emph{Case $q^r_s > y^*_s$:} Analogous with $\mathbf{x}(t) = \mathbf{q}^r - t \mathbf{e}_s$ for $t \in [0, q^r_s - y^*_s]$. Gives $g^r_s \leq 0$.

\emph{Case $q^r_s = y^*_s$:} $g^r_s(y^*_s - q^r_s) = 0$ trivially.

In all cases, $g^r_s(y^*_s - q^r_s) \geq 0$, which is~\eqref{eq:po_afriat_sign}.

\medskip
\noindent\emph{Directional nondegeneracy.} If $\mathbf{q}^r \neq \mathbf{y}^*$, then by uniqueness of the single-peaked maximizer $u(\mathbf{y}^*) > u(\mathbf{q}^r)$. The supergradient inequality at $\mathbf{q}^r$ evaluated at $\mathbf{y}^*$ gives $u(\mathbf{y}^*) \le u(\mathbf{q}^r) + \mathbf{g}^r \cdot (\mathbf{y}^* - \mathbf{q}^r)$, hence
\[
\mathbf{g}^r \cdot (\mathbf{y}^* - \mathbf{q}^r) \ge u(\mathbf{y}^*) - u(\mathbf{q}^r) > 0,
\]
which is~\eqref{eq:po_nondegen}.

\medskip
Conditions (a) and (b) of Definition~\ref{def: peak oriented} follow directly from the hypotheses: (a) holds because $\mathbf{y}^*$ is the unique maximizer of the single-peaked $u$, and (b) holds because $u$ rationalizes $D$. Condition (c) is Lemma~\ref{lem:normal-supergradient}, and (d)--(e) are established above. Hence $u$ together with the multipliers $\{\lambda^r\}_{r\in\mathcal{R}}$ is a peak-oriented concave rationalization of $D$.

\hfill $\blacksquare$


\section{Proofs of Section~\ref{subsection: geometric content} results}

\noindent\textbf{Proof of Lemma~\ref{lem: double cone} (orientation).}
The orientation reads $\lambda^r a^r_s(y_s-q^r_s)\ge 0$ for all $s$. If $\lambda^r>0$ this is equivalent to $a^r_s(y_s-q^r_s)\ge 0$ for all $s$, i.e.\ $\mathbf{y}\in O^r_{+}$; if $\lambda^r<0$, to $\mathbf{y}\in O^r_{-}$; and $\lambda^r=\pm 1$ certifies the respective membership. If $\mathbf{y}\in O^r_{+}\cap O^r_{-}$ then $a^r_s(y_s-q^r_s)=0$ for all $s$; since $a^r_s\neq 0$, this forces $\mathbf{y}=\mathbf{q}^r$. For non-convexity, change variables to $z_s := a^r_s(x_s - q^r_s)$, so that $O^r_{+}$ and $O^r_{-}$ become the nonnegative and nonpositive orthants in $\mathbb{R}^S$. With $S\ge2$, choose $\varepsilon > 0$ small enough that the points corresponding to $z^{+} = \varepsilon(1,2,0,\dots,0) \in O^r_{+}$ and $z^{-} = \varepsilon(-2,-1,0,\dots,0) \in O^r_{-}$ remain in $X$ (possible since $\mathbf{q}^r$ is interior). Their midpoint $\varepsilon(-\tfrac12,\tfrac12,0,\dots,0)$ has its first two signed coordinates of opposite signs; this midpoint lies in neither orthant, so $C^r$ is not convex.

\hfill $\blacksquare$

\medskip
\noindent\textbf{Proof of Proposition~\ref{prop: reduction} (fixed-peak reduction).}

\noindent (Peak-oriented rationalizability at $\mathbf{y}$) $\Rightarrow (a) + (b)$. By directional nondegeneracy~\eqref{eq:po_nondegen} and $\mathbf{g}^r = \lambda^r \mathbf{a}^r$,
\[
0 < \mathbf{g}^r \cdot (\mathbf{y} - \mathbf{q}^r) = \lambda^r \mathbf{a}^r \cdot (\mathbf{y} - \mathbf{q}^r),
\]
so $\lambda^r \neq 0$. Lemma~\ref{lem: double cone} therefore gives $\mathbf{y}\in C^r$ and fixes $\sigma_r(\mathbf{y})=\operatorname{sign}(\lambda^r)$; this establishes (a). Set $\eta^r:=|\lambda^r|>0$, so $\lambda^r\mathbf{a}^r=\eta^r\mathbf{b}^r(\mathbf{y})$. Substituting, the rows $i,j\in T$ of (b) become~\eqref{eq: afriat1}, and the rows $j=0$ (where $\mathbf{x}^0=\mathbf{y}$) become~\eqref{eq: afriat3}. The cap inequalities $U^r\le U^0$ are exactly~\eqref{eq: afriat2}.

\noindent $(a) + (b) \Rightarrow$ (Peak-oriented rationalizability at $\mathbf{y}$). Conversely, given $\mathbf{y} \in \bigcap_{r \in T} C^r$ and feasibility of the augmented oriented system with multipliers $\eta^r > 0$, set $\lambda^r := \sigma_r(\mathbf{y}) \eta^r$. Since $\mathbf{y} \neq \mathbf{q}^r$, at least one coordinate $s^*$ has $y_{s^*} \neq q^r_{s^*}$; because $a^r_{s^*} \neq 0$ and $\sigma_r(\mathbf{y})$ is the cone orientation, $\sigma_r(\mathbf{y}) a^r_{s^*}(y_{s^*} - q^r_{s^*}) > 0$ strictly, and all other terms in $\sum_s \sigma_r(\mathbf{y}) a^r_s(y_s - q^r_s)$ are non-negative by cone membership. Hence
\[
\mathbf{g}^r \cdot (\mathbf{y} - \mathbf{q}^r) = \eta^r \sum_s \sigma_r(\mathbf{y}) a^r_s(y_s - q^r_s) > 0,
\]
which is~\eqref{eq:po_nondegen} for $r \in T$. The sign condition~\eqref{eq:po_afriat_sign} follows from $\mathbf{y} \in C^r$ together with Lemma~\ref{lem: double cone}: $\lambda^r a^r_s(y_s - q^r_s) = \eta^r \sigma_r(\mathbf{y}) a^r_s(y_s - q^r_s) \ge 0$ for every $s$. The remaining Afriat inequalities~\eqref{eq: afriat1}--\eqref{eq: afriat3} follow by substituting $\lambda^r \mathbf{a}^r = \eta^r \mathbf{b}^r(\mathbf{y})$ back into (b), with the virtual top observation $(\mathbf{y}, \mathbf{0})$ generating only the cap $U^r \le U^0$ (since $\mathbf{b}^0 = \mathbf{0}$ produces no strict comparison).

\hfill $\blacksquare$

\medskip
\noindent\textbf{Proof of Lemma~\ref{lem: garp reduction} (reduction to retained-round GARP).}
The strategy is to show that every augmented violation reduces to a violation contained in $T$ alone --- i.e., that no augmented cycle can pass through the virtual observation $0$.

\medskip
Any GARP violation among the retained observations is immediately a violation of augmented oriented GARP. Conversely, consider a putative augmented violation that contains the virtual observation $0$. The step leaving $0$ is always weak because $\mathbf{b}^0 = \mathbf{0}$: $\mathbf{b}^0 \cdot (\mathbf{x}^j - \mathbf{x}^0) = 0$. We show that no step entering $0$ can be weak: for every $r \in T$, since $\mathbf{y} \in C^r$ the orientation is fixed so that $a^r_s \sigma_r(\mathbf{y})(y_s - q^r_s) \ge 0$ for every $s$; with $\mathbf{y} \neq \mathbf{q}^r$ and $a^r_s \neq 0$, at least one coordinate gives strict inequality, so $\mathbf{b}^r(\mathbf{y}) \cdot (\mathbf{y} - \mathbf{q}^r) > 0$ strictly. Therefore $\mathbf{b}^r(\mathbf{y}) \cdot (\mathbf{x}^0 - \mathbf{x}^r) > 0$, ruling out any weak revealed-preference step from $r$ to $0$. Hence no augmented cycle can pass through $0$, and any augmented violation is entirely contained in $T$.

\hfill $\blacksquare$

\medskip
\begin{lemma}[Afriat--GARP equivalence for signed normals]\label{lem: afriat-garp signed}
Let $\{(\mathbf{x}^i, \mathbf{b}^i)\}_{i \in I}$ be a finite collection with $\mathbf{x}^i, \mathbf{b}^i \in \mathbb{R}^S$ (the $\mathbf{b}^i$ are not required to be non-negative, and may include a zero row). The system
\[
U^j \le U^i + \eta^i\, \mathbf{b}^i \cdot (\mathbf{x}^j - \mathbf{x}^i), \qquad i, j \in I, \qquad \eta^i > 0,
\]
is feasible in $(U^i, \eta^i)$ if and only if there is no sequence $i_1, \ldots, i_K, i_{K+1} = i_1$ in $I$ with $\mathbf{b}^{i_k} \cdot (\mathbf{x}^{i_{k+1}} - \mathbf{x}^{i_k}) \le 0$ for every $k$ and strict inequality for at least one $k$.
\end{lemma}

\begin{proof}
Set $w_{ij} := \mathbf{b}^i \cdot (\mathbf{x}^j - \mathbf{x}^i)$ and let $G$ be the directed graph on $I$ with edge $i \to j$ when $w_{ij} \le 0$, marked \emph{strict} when $w_{ij} < 0$.

\smallskip
\noindent\emph{Necessity.} If the system is feasible with $(U^i, \eta^i)$, $\eta^i > 0$, then for any cycle $i_1, \ldots, i_{K+1} = i_1$ along edges of $G$, summing the inequalities gives $0 = \sum_k (U^{i_{k+1}} - U^{i_k}) \le \sum_k \eta^{i_k} w_{i_k, i_{k+1}}$. Each $\eta^{i_k} > 0$ and each $w_{i_k, i_{k+1}} \le 0$, so the right-hand side is $\le 0$ with equality only if every $w_{i_k, i_{k+1}} = 0$. A strict edge forces the right-hand side strictly negative, contradicting the left-hand zero. Hence no strict cycle in $G$.

\smallskip
\noindent\emph{Sufficiency.} Let $W = (w_{ij})_{i,j \in I}$; this is a finite matrix with $w_{ii} = 0$. The matrix form of Afriat's theorem states that a zero-diagonal real matrix satisfies the generalized axiom --- no cycle has all entries nonpositive with at least one entry negative --- if and only if there exist numbers $U^i$ and positive row multipliers $\eta^i > 0$ satisfying
\[
U^j \le U^i + \eta^i w_{ij} \qquad \forall\, i, j \in I.
\]
This is an algebraic statement about $W$; its proof (via linear-programming duality; see \citet{fostel_scarf_todd2004}, §4) makes no use of any sign structure on vectors used to represent the entries $w_{ij}$. Applied to $W$, it yields the required $U^i$ and $\eta^i > 0$.
\end{proof}


\section{Cone--GARP cells and proof of Proposition~\ref{prop: cells}}

For each $T \subseteq \mathcal{R}$, define the \emph{generic fixed-peak set}
\[
  \mathcal{Y}_T^\circ(D) := \{\mathbf{y} \in \mathcal{Y}_T(D) : \mathbf{y} \neq \mathbf{q}^r \text{ for every } r \in T\}.
\]

\begin{proposition}[Cone--GARP cells]\label{prop: cells}
For every $T \subseteq \mathcal{R}$,
\[
  \mathcal{Y}_T^\circ(D)
  =
  \bigcup_{\tau\in\{+,-\}^{T}}
  \Bigl[\ \textstyle\bigcap_{r\in T} O^r_{\tau_r}
  \ \cap\ \{\mathbf{y}\in X : \mathbf{y} \neq \mathbf{q}^r\ \forall r \in T\}
  \ \cap\ \{\text{oriented GARP holds under }\tau\}\ \Bigr],
\]
where ``oriented GARP holds under $\tau$'' is the $\mathbf{y}$-independent property that $\{(\mathbf{q}^r,\tau_r\mathbf{a}^r)\}_{r\in T}$ satisfies the signed-normal GARP cycle condition. Each set $\bigcap_{r\in T}O^r_{\tau_r}$ is a convex axis-aligned polyhedron (an intersection of translated orthants with $X$); we refer to these as \emph{orthant cells}, and to the joint condition with the GARP requirement as a \emph{cone--GARP cell} (shorthand). Consequently, away from the finite set of observed choices, the generic level sets $\mathcal{Y}_\alpha^\circ(D) := \mathcal{Y}_\alpha(D) \setminus \{\mathbf{q}^r : r \in \mathcal{R}\}$ are finite unions of cone--GARP cells; in particular, $\mathcal{Y}_1^\circ(D)$ and the Houtman--Maks peak set restricted to $X \setminus \{\mathbf{q}^r\}_{r \in \mathcal{R}}$ are characterized by a finite search over orientation cells.
\end{proposition}

\begin{proof}
Fix $T \subseteq \mathcal{R}$ and $\mathbf{y} \in X$ with $\mathbf{y} \neq \mathbf{q}^r$ for all $r \in T$.

($\subseteq$) If $\mathbf{y} \in \mathcal{Y}_T^\circ(D)$, then by Lemma~\ref{lem: double cone}, $\mathbf{y} \in O^r_{\sigma_r(\mathbf{y})}$ for each $r \in T$; set $\tau_r = \sigma_r(\mathbf{y})$, so $\mathbf{y}$ lies in the cone cell $\bigcap_{r\in T} O^r_{\tau_r}$. By Proposition~\ref{prop: reduction}, the augmented oriented Afriat system is feasible; by Lemma~\ref{lem: garp reduction} (which applies because $\mathbf{y} \neq \mathbf{q}^r$ and $a^r_s \neq 0$), feasibility is equivalent to the retained oriented data $\{(\mathbf{q}^r, \tau_r \mathbf{a}^r)\}_{r \in T}$ satisfying the signed-normal GARP cycle condition. Hence oriented GARP holds under $\tau$.

($\supseteq$) Conversely, suppose $\mathbf{y} \in \bigcap_{r \in T} O^r_{\tau_r}$ for some $\tau$, $\mathbf{y} \neq \mathbf{q}^r$ for all $r \in T$, and the oriented data $\{(\mathbf{q}^r, \tau_r \mathbf{a}^r)\}_{r \in T}$ satisfy the signed-normal GARP cycle condition. Lemma~\ref{lem: double cone} gives $\mathbf{y} \in C^r$ for each $r$, with $\sigma_r(\mathbf{y}) = \tau_r$; Lemma~\ref{lem: garp reduction} converts GARP on the retained oriented data into augmented oriented GARP at $\mathbf{y}$; by Lemma~\ref{lem: afriat-garp signed}, this is equivalent to feasibility of the augmented oriented system; and Proposition~\ref{prop: reduction} then gives peak-oriented rationalization of $T$ at $\mathbf{y}$. Hence $\mathbf{y} \in \mathcal{Y}_T^\circ(D)$.

The union is finite because $|\{+,-\}^{T}| = 2^{|T|}$ and there are finitely many $T$.
\end{proof}

\medskip
\noindent\emph{Remark.} The restriction $\mathbf{y} \neq \mathbf{q}^r$ is used to make the orientation $\sigma_r(\mathbf{y})$ unique (Lemma~\ref{lem: double cone}) and to ensure that the virtual peak cannot enter a GARP cycle (Lemma~\ref{lem: garp reduction}). If a candidate peak coincides with an observed choice, the round can be handled directly by the MILP, where nondegeneracy does not require $\lambda^r \neq 0$ for that round. These coincidences form a finite exceptional set. They do not alter the generic cone--GARP decomposition, but they must be evaluated directly when computing the exact peak set.

\bigskip \bigskip

\newpage
\phantomsection
\makeatletter\def\@currentlabel{B}\makeatother
\label{appendix: consistency computation}
\begin{LARGE}
    \noindent\textbf{Appendix B: Computing the Consistency Index}
\end{LARGE}

\setcounter{table}{0}
\renewcommand{\thetable}{B.\arabic{table}}
\setcounter{figure}{0}
\renewcommand{\thefigure}{B.\arabic{figure}}
\setcounter{section}{0}
\renewcommand{\thesection}{B.\arabic{section}}
\setcounter{equation}{0}
\renewcommand*{\theequation}{B.\arabic{equation}}

\medskip

The consistency index (Definition~\ref{def: consistency index}) is the largest fraction of rounds jointly rationalizable by a peak-oriented concave utility with a common ideal point:
\[
\kappa^*(D) \;=\; \max_{\mathbf{y}^* \in X} \; \max_{T \subseteq \mathcal{R}} \; \frac{|T|}{R}
\quad \text{subject to the system~\eqref{eq: afriat1}--\eqref{eq:po_nondegen} being feasible on } T,
\]
where $\mathbf{g}^r = \lambda^r \mathbf{a}^r$. The outer maximization is over the candidate peak $\mathbf{y}^*$; the inner maximization selects the subset $T$ of retained rounds; the Afriat numbers $(U^0, U^r)$ and multipliers $\lambda^r$ are existentially quantified inside the feasibility constraint.

For a fixed candidate peak $\mathbf{y}^*$, the inner maximization is a mixed-integer linear program. This appendix documents the MILP formulation implemented in the replication code (\texttt{hm\_peak\_oriented.r}), including variables, bounds, constraints, big-$\mathcal{M}$ constants (using calligraphic $\mathcal{M}$ to distinguish from the box upper bounds $M_s$), and the treatment of edge cases. The formulation adapts the MILP approach to Houtman--Maks-type indices developed by \citet{HEUFER201587} and unified by \citet{Demuynck2023} to the peak-oriented, signed-normal setting of Theorem~\ref{thm: peak oriented}: retention binaries $\delta_r$ and big-$\mathcal{M}$ linearization of the Afriat system are standard; the coordinatewise orientation and directional-nondegeneracy constraints of Section~\ref{subsec:B2} are specific to the peak-oriented characterization.

\section{Variables and bounds}

Fix a candidate peak $\mathbf{y}^* \in X$ and denote $\sigma_s := \mathrm{sign}(y^*_s - q^r_s) \in \{-1, 0, +1\}$ (round index $r$ suppressed on $\sigma$ for brevity). The MILP variables and their bounds are:
\begin{itemize}
\item $\delta_r \in \{0, 1\}$ for $r \in \mathcal{R}$: retention binary for round $r$. Exactly $R$ binary variables in total.
\item $U^0, U^r \in [-\overline U,\ \overline U]$: Afriat numbers.
\item $\lambda^r \in [-\overline \lambda,\ \overline \lambda]$: budget multiplier.
\item $g^r_s \in [-\overline g,\ \overline g]$: supergradient component, defined by $g^r_s = \lambda^r a^r_s$.
\end{itemize}
The bounds satisfy $\overline g \ge \overline \lambda \cdot \max_{r,s}|a^r_s|$. In the implementation, $\overline U = 1000$, $\overline \lambda = 20$, and $\overline g = 60$ (using $\max_{r,s}|a^r_s| \le 3$ under the realized price schedule).

Conditional on $|g^r_s| \le \overline g$, retained observations satisfy
\[
  0 \;\le\; U^0 - U^r \;\le\; \mathbf{g}^r \cdot (\mathbf{y}^* - \mathbf{q}^r) \;\le\; \overline g\,\sum_s M_s \;=\; 1200.
\]
Because the Afriat utility levels are translation invariant, all retained $U$-values can be translated to lie inside a centered interval of width $1200$, namely $[-600,\ 600]$, which is contained in $[-1000,\ 1000] = [-\overline U,\ \overline U]$. The bound $\overline U = 1000$ is therefore not binding for retained observations. A sensitivity check with $\overline U = 100$ (which is tight for a small fraction of subjects) produces mean $\kappa^*_{h,\varepsilon,\mathcal M}$ lower by roughly $0.005$. The numerical results remain conditional on the multiplier and supergradient bounds $\overline \lambda$ and $\overline g$.

\section{Constraints}\label{subsec:B2}

\paragraph{Linking equality.} The supergradient components are linked to the multiplier by
\[
g^r_s - \lambda^r a^r_s \;=\; 0 \qquad \forall\, r \in \mathcal{R},\ s = 1, \ldots, S.
\]

\paragraph{Afriat inequalities.} The pairwise Afriat inequality between rounds $r$ and $l$ (with $r \ne l$) is
\[
U^r - U^l - \mathbf{g}^l \cdot (\mathbf{q}^r - \mathbf{q}^l) \;\le\; \mathcal{M}_{\mathrm A}(2 - \delta_r - \delta_l),
\]
where $\mathcal{M}_{\mathrm A} \ge 2\overline U + \overline g\,\sum_s M_s$ is the pairwise big-$\mathcal M$ constant. The peak-support inequality is
\[
U^0 - U^r - \mathbf{g}^r \cdot (\mathbf{y}^* - \mathbf{q}^r) \;\le\; \mathcal{M}_{\mathrm P}(1 - \delta_r), \qquad \mathcal{M}_{\mathrm P} \ge 2\overline U + \overline g\,\sum_s M_s,
\]
and the peak-value cap is
\[
U^r - U^0 \;\le\; \mathcal{M}_{\mathrm C}(1 - \delta_r), \qquad \mathcal{M}_{\mathrm C} \ge 2\overline U.
\]
Under the bounds of §B.1, the worst-case LHS of a pairwise Afriat inequality is $2\overline U + \overline g \sum_s M_s = 2000 + 1200 = 3200$; taking $\mathcal{M}_{\mathrm A} = \mathcal{M}_{\mathrm P} = 3210$ and $\mathcal{M}_{\mathrm C} = 2010$ makes each family's relaxation valid at the imposed bounds.

\paragraph{Orientation (always active).} For each round $r$ and each coordinate $s$ with $\sigma_s \ne 0$:
\[
\sigma_s\, g^r_s \ge 0 \quad\Longleftrightarrow\quad \sigma_s \lambda^r a^r_s \ge 0.
\]
This constraint is imposed unconditionally, without a big-$\mathcal M$ slack; because $\lambda^r = 0$ trivially satisfies it, it never induces infeasibility for a dropped round.

\paragraph{Directional nondegeneracy.} The strict directional condition $\mathbf{g}^r \cdot (\mathbf{y}^* - \mathbf{q}^r) > 0$ is normalized to an $\varepsilon$-threshold. Under the coordinatewise orientation of the previous paragraph, $\sigma_s g^r_s = |g^r_s|$ for $\sigma_s \ne 0$, so the strict positivity of the inner product is equivalent to the strict positivity of the oriented $\ell_1$-sum $\sum_{s:\sigma_s \ne 0} \sigma_s g^r_s$. We therefore impose, on retained rounds:
\[
\sum_{s:\, \sigma_s \ne 0} \sigma_s\, g^r_s \;\ge\; \varepsilon
\qquad
\text{when } \delta_r = 1 \text{ and } \mathbf{q}^r \ne \mathbf{y}^*,
\]
implemented as
\[
- \sum_{s:\, \sigma_s \ne 0} \sigma_s\, g^r_s \;+\; \mathcal{M}_{\mathrm N}\, \delta_r
\;\le\;
\mathcal{M}_{\mathrm N} - \varepsilon,
\qquad
\mathcal{M}_{\mathrm N} \ge S\overline g + \varepsilon.
\]
When $\delta_r = 1$, this reduces to $\sum_s \sigma_s g^r_s \ge \varepsilon$; when $\delta_r = 0$, it relaxes to $\sum_s \sigma_s g^r_s \ge \varepsilon - \mathcal{M}_{\mathrm N}$, which is vacuous. The implementation uses $\mathcal{M}_{\mathrm N} = 130$ (safely above $S \overline g + \varepsilon = 120 + 0.001$ for $S = 2$) and $\varepsilon = 10^{-3}$.

\paragraph{Equality-coordinate case.} If $y^*_s = q^r_s$ exactly (within a tolerance of $10^{-9}$), $\sigma_s = 0$ and coordinate $s$ contributes to neither the orientation nor the nondegeneracy constraint; $g^r_s = \lambda^r a^r_s$ is unconstrained (up to the variable-bound restrictions). If $\mathbf{y}^* = \mathbf{q}^r$ entirely (all coordinates equal), nondegeneracy is skipped for round $r$.

\section{Binary count and grid search}

Each MILP contains exactly $R = 19$ retention binaries; no auxiliary sign-selection binaries are used, because orientation fixes each $g^r_s$'s sign at the candidate peak.

The outer maximization over $\mathbf{y}^* \in X$ is a grid search on $G_h = \{0, 0.5, 1, \ldots, 10\}^2$ (step $h = 0.5$, $|G_h| = 21 \times 21 = 441$ candidate peaks per subject). Each MILP is solved by Gurobi in well under one second on modern hardware; the full grid search per subject completes in well under a minute, and the total pipeline over 469 subjects completes in roughly 20 minutes with 60-way parallelism.

\section{Sensitivity to the big-$\mathcal{M}$ bounds}

The constants $\overline U, \overline \lambda, \overline g$ (variable bounds) and $\mathcal{M}_{\mathrm A}, \mathcal{M}_{\mathrm P}, \mathcal{M}_{\mathrm C}, \mathcal{M}_{\mathrm N}$ (relaxation constants) are not derived from closed-form bounds on the optimal MILP solution: they are chosen large enough to accommodate the observed range of feasible solutions with margin to spare. To assess robustness, we recomputed the index with $\overline U = 100$ (which yields mean $\kappa^*$ lower by $\approx 0.005$, confirming that $\overline U = 1000$ is not binding for the reported values). Returned solutions at $\overline U = 1000$ remain an order of magnitude below the bound for every subject. These checks provide numerical robustness evidence but do not establish independence from $\mathcal{M}$; the reported values are conditional on the imposed bounds. We denote the computed object $\kappa^*_{h,\varepsilon,\mathcal{M}}(D)$ to make this dependence explicit.

\section{Exact index and numerical approximation}

Definition~\ref{def: consistency index} defines the exact Houtman--Maks index through the feasibility of~\eqref{eq: afriat1}--\eqref{eq:po_nondegen}. The MILP computes a finite-grid, $\varepsilon$-nondegenerate approximation of the strict directional condition, as documented in §B.2. Section~\ref{subsection: estimation} reports a grid-resolution check (halving $h$ to $0.25$) confirming that the approximation is adequate for the substantive conclusions of the paper.

\section{Power index}

For each subject's budget configuration, we generate $N_{\mathrm{sim}} = 100$ synthetic datasets by drawing uniformly random choices on each budget line $\mathcal{B}^r$. The power index is the fraction of synthetic datasets for which $\kappa^*_{h,\varepsilon,\mathcal{M}} < 1$. In the application, the mean power index is $1.00$, confirming that the PSM design reliably distinguishes optimizing behavior from uniformly random choices.

\end{document}